\documentclass{aastex62}
\received{ }
\revised{ }
\accepted{ }
\submitjournal{ApJS}

\shorttitle{Sample article}
\shortauthors{Luo et al.}

\begin{document}

\title{Hot subdwarf stars observed in Gaia DR2 and LAMOST DR5}

\correspondingauthor{Yangping Luo}
\email{ypluo@bao.ac.cn}

\author{Yangping Luo}
\affiliation{Department of Astronomy, China West Normal University, \\
Nanchong, 637002, PR China}

\author{P\'{e}ter N\'{e}meth}
\affiliation{Astronomical Institute of the Czech Academy of Sciences, CZ-25\,165, Ond\u{r}ejov, Czech Republic}
\affiliation{Astroserver.org, 8533 Malomsok, Hungary}

\author{Licai Deng}
\affiliation{Department of Astronomy, China West Normal University, \\
Nanchong, 637002, PR China}
\affiliation{Key Laboratory of Optical Astronomy, National Astronomical Observatories, Chinese Academy of Sciences, Beijing 100012, China}

\author{Zhanwen Han}
\affiliation{Key Laboratory for the Structure and Evolution of Celestial Objects, Chinese Academy of Sciences, \\
Kunming 650011, PR China}



\begin{abstract}
Combing Gaia DR2 with LAMOST DR5, we spectroscopically identified 924 hot subdwarf stars, among which 32 stars exhibit strong double-lined composite spectra.
We measured the effective temperature $T_{\rm eff}$, surface gravity $\log\,g$, helium abundance $y=n{\rm He}/n{\rm H}$, and radial velocities of 892 non-composite spectra hot subdwarf stars by fitting LAMOST observations with {\sc Tlusty/Synspec} non-LTE synthetic spectra. We outlined four different groups in the $T_{\rm eff}-\log\,g$ diagram with our helium abundance classification scheme and two nearly parallel sequences in the $T_{\rm eff}-\log(y)$ diagram. 3D Galactic space motions and orbits of 747 hot subdwarf stars with $(G_{BP}-G_{RP})_{0}<-0.36$ mag were computed  using LAMOST radial velocities and Gaia parallaxes and proper motions. Based on the $U-V$ velocity diagram, $J_{z}-$eccentricity diagram, and Galactic orbits, we derived Galactic population classifications and the fractional distributions of the four hot subdwarf helium groups in the halo, thin disk and thick disk. Comparisons with the predictions of binary population synthesis calculations \citep{2008A&A...484L..31H}
suggest that He-rich hot subdwarf stars with $\log(y)\ge0$ are from the double helium white dwarfs merger, He-deficient hot subdwarf stars with $-2.2\le\log(y)<-1$ from the common envelope ejection, and He-deficient hot subdwarf stars with $\log(y)<-2.2$ from the stable Roche lobe overflow channels. The relative number of He-rich hot subdwarf stars with $-1\le\log(y)<0$ and $\log(y)\ge0$ in the halo is more than twice the prediction of \cite{2017ApJ...835..242Z}, even more than six times in the thin disk,
which implies that the mergers of helium white dwarfs with low mass
main sequence stars may not be the main formation channel of He-rich hot subdwarf stars with $-1\le\log(y)<0$, specially in younger environments.
\end{abstract}


\section{Introduction} \label{sec:intro}
Hot subdwarf stars are core helium-burning stars. Their optical spectra resemble that of O and B main sequence (MS) stars, but their typical luminosity is much lower than on the MS. In the
Hertzsprung-Russell (HR) diagram they are located at the blue end of the horizontal branch (HB) below O and B type MS, also known as the extreme horizontal
branch (EHB). In general, these stars were traditionally divided into O type subdwarf (sdO) and B type subdwarf (sdB) stars based on their spectral features \citep{2013A&A...551A..31D}. They were
discovered in all Galactic stellar populations \citep{2009ARA&A..47..211H,2016PASP..128h2001H} and turned out to be important objects to account for
the $UV$ excess in the spectra of elliptical galaxies and the bulges of spiral galaxies \citep{2007MNRAS.380.1098H}, and they play an important role to understand the horizontal branch
morphology of globular clusters \citep{2008A&A...484L..31H, 2013A&A...549A.145L, 2015MNRAS.449.2741L}. They also play a key role in stellar astrophysics.
Most hot subdwarf stars display large anomalies of helium abundances, which may be an ideal trace element to explore diffusion processes in stellar atmospheres
\citep{2003A&A...400..939E, 2016PASP..128h2001H, 2018MNRAS.475.4728B}. Moreover, a few helium-rich hot subdwarf stars exhibit extreme surface chemical
compositions, with abundances of lead, zirconium, strontium, and yttrium up to $10\,000$ times the solar value \citep{2011MNRAS.412..363N,2013MNRAS.434.1920N,2017MNRAS.465.3101J}.
Several classes of pulsating stars have been discovered among hot subdwarfs, so that asteroseismology has become an important tool to probe their interior
structure \citep{2012A&A...539A..12F, 2014A&A...569A..15O, 2018ApJ...853...98Z,2019MNRAS.482..758S}. Hot subdwarf binaries with massive CO white dwarfs have been
proposed as a formation channel of type Ia Supernovae \citep{2009A&A...493.1081J, 2009MNRAS.395..847W, 2010A&A...515A..88W, 2013A&A...554A..54G, 2015Sci...347.1126G, 2018RAA....18...49W}.
Binary systems consisting of hot subdwarf stars and neutron stars are potential gravitational wave sources that might be resolved by the Laser
Interferometer Space Antenna (LISA) in the future \citep{2018arXiv180803402W}. More details on hot subdwarf stars have been given by \cite{2009ARA&A..47..211H} and recent developments in this rapidly evolving field have been reviewed by \cite{2016PASP..128h2001H}.

However, the formation of hot subdwarf stars is still wrapped in mystery. The main difficulty is how to lose almost the complete hydrogen envelope prior to
or at the beginning of the helium core flash. A number of scenarios trying to interpret the formation of hot subdwarf stars have been put forward.
Due to the findings of the high fraction of binaries amongst hot subdwarfs, interacting binary evolution through common envelope (CE) ejection, stable Roche
lobe overflow (RLOF)  and the merger of double helium white dwarfs (HeWD) were proposed to be the main formation channels \citep{1984ApJ...277..355W,2002MNRAS.336..449H,2003MNRAS.341..669H}.
The first two channels are responsible mainly for sdB stars and the last one for He-rich sdO stars \citep{2002MNRAS.336..449H, 2008A&A...484L..31H, 2012MNRAS.419..452Z}.
Both the late hot-flasher scenario \citep{1996ApJ...466..359D,2004A&A...415..313M,2008A&A...491..253M} and the mergers of helium white dwarfs with low mass main
squence stars \citep{2017ApJ...835..242Z} provide further options to explain some observed intermediate helium-rich hot subdwarf stars. Although these models
can explain the current observations of hot subdwarfs, none of them appears entirely satisfactory. The main reason is that some crucial physical processes
(mass loss on the red-giant branch, surface element diffusion, common-envelope evolution, mass transfer, etc.) cannot be dealt with the desired accuracy
\citep{2016PASP..128h2001H}. On the other hand, the currently available observations \citep{2003A&A...400..939E, 2005A&A...430..223L, 2007A&A...462..269S, 2009PhDT.......273H, 2012MNRAS.427.2180N, 2011A&A...530A..28G, 2015A&A...577A..26G, 2016ApJ...818..202L, 2018ApJ...868...70L} cannot provide enough information to constrain these models.
Therefore, kinematic investigations and spectral analyses on large and homogeneous samples are very valuable to understand the formation of
hot subdwarf stars.

LAMOST (the Large Sky Area Multi-Object Fiber Spectroscopic Telescope, also named the "Guo Shou Jing" Telescope) is a 4-m specially designed Schmidt survey
telescope located at the Xinglong Station of the National Astronomical Observatories of the Chinese Academy of Sciences, which can simultaneously take the spectra
of 4000 objects in a field of view of about $5^{\circ}$ in diameter \citep{2012RAA....12.1197C}. Up to December of 2017, the LAMOST survey has obtained more than
9 million spectra, which were internally released as Data Release 5 (DR5; see \url{http://dr5.lamost.org}). A total of 166 hot subdwarf stars were identified from the LAMOST DR1 spectra
by \cite{2016ApJ...818..202L}. Gaia, as the astrometric successor of the Hipparcos mission of the European Space Agency (ESA), is an ambitious space mission designed
for measuring highly accurate positions, parallaxes and proper motions of about one billion stars in the Milky Way. Gaia DR2 is setting a major step in finding hot subdwarf stars, because it provides more than 1.3 billion sources with high precision positions, trigonometric parallaxes, and proper motions as well as three
broad band magnitudes ($G$, $G_{\rm BP}$, and $G_{\rm RP}$) \citep{2018A&A...616A...1G,2018A&A...616A..10G,2018A&A...616A..11G}. \cite{2019A&A...621A..38G} published
an all-sky catalogue of $39\,800$ hot subdwarf candidates selected from Gaia DR2. \cite{2018ApJ...868...70L} spectroscopically confirmed 294 new hot subdwarf stars
in Gaia DR2 with LAMOST DR5 spectra. The combination of LAMOST spectra with Gaia data provides a good opportunity to reveal the origins of the different types of hot
subdwarf stars because some new observational constraints can be outlined from kinematics and spectral analyses.

This paper presents the spectral analyses of 892 non-composite spectra hot subdwarf stars and the kinematics of 747 stars with LAMOST DR5 spectra and Gaia DR2.
In Section$\,$\ref{sec:data}, we describe LAMOST DR5, GAIA DR2 and our sample selection. Section$\,$\ref{sec:atmo} gives the determination of atmospheric parameters and
the physical properties of hot subdwarf stars. Section$\,$\ref{sec:spac} and$\,$\ref{sec:kine} present the Galactic space distributions and kinematics of all targets. Galactic orbits are computed in Section$\,$\ref{sec:orb}. Galactic population classifications of the different helium abundance hot subdwarf stars are presented in Section$\,$\ref{sec:pop} and conclusions are drawn in Section$\,$\ref{sec:conc}.

\section{Data and Sample selection} \label{sec:data}
\subsection{Data}
LAMOST DR5 was released to the public in December of 2017 and published more than 9 million spectra, including $8\,171\,443$ stars, $153\,090$ galaxies,
$51\,133$ quasars, and $642\,178$ unclassified objects. LAMOST DR5 also provided the effective temperatures, surface gravities, and metallicities of $5\,344\,058$ stars.
The target selection algorithm and strategy of observations of the LAMOST Spectroscopic Survey can be found in \cite{2012RAA....12..755C, 2012RAA....12..805C, 2012RAA....12..792Z, 2015MNRAS.448..822X, 2015MNRAS.448..855Y}.
The LAMOST spectra are similar to the SDSS data having the resolution $R\thicksim1800$ and wavelength coverage of $3800-9100$ \AA. Three data processing pipelines were developed by the LAMOST data center \citep{2012RAA....12.1243L, 2014IAUS..298..428L}.
The details have been described in \cite{2016ApJ...818..202L}.

Gaia DR2 was released on 25 April 2018 and provided high-precision positions, proper motions, and trigonometric parallaxes as well as three broad-band magnitudes
($G$, $G_{\rm BP}$, and $G_{\rm RP}$) for more than 1.3 billion objects with a limiting magnitude of $G=20$ and a bright limit of $G=3$ \citep{2018A&A...616A...1G,2018A&A...616A..10G, 2018A&A...616A..11G}.
Parallax uncertainties are up to 0.04 milliarcseconds (mas) for sources with $G<15$, around 0.1$\,$mas for sources at $G=17$, and around 0.7$\,$mas for sources at $G=20$.
The corresponding uncertainties in proper motions are in the range of $0.06\,{\rm mas}\,{\rm yr}^{-1}$ for $G<15$, $0.2\,{\rm mas}\,{\rm yr}^{-1}$ for $G=17$, and $1.2\,{\rm mas}\,{\rm yr}^{-1}$ for
$G=20$. For a detailed description of the Gaia DR2 data, see \cite{2018A&A...616A..11G,2018A&A...616A..12G}.

\subsection{Sample selection}\label{subsec:samp}
We utilized the Gaia HR diagram to select targets. Hot subdwarf candidates can be identified visually in the Gaia HR diagram \citep{2018ApJ...868...70L}.
There are more than 8 million common sources by cross-matching LAMOST DR5 and Gaia DR2.
Due to the effects of extinction, there are many hot subdwarfs in the overlap region with O and B type MS stars.
In order to plot de-reddened HRD, Gaia color $G_{\rm BP}-G_{\rm RP}$ and magnitude $G$ were corrected by following the method of \cite{2018ApJ...866..121H}.
Firstly, we obtained the $B-V$ color excess $E(B-V)$ from the dust maps of \citet{1998ApJ...500..525S} and then
estimated the $V-I$ color excess $E(V-I)$ with the relation of $E(V-I)=1.18E(B-V)$ \citep{2011ApJ...737..103S}.
Finally, the Gaia color excess $E(G_{\rm BP}-G_{\rm RP})$ was inferred from the relationship:
\begin{equation}
E(G_{\rm BP}-G_{\rm RP})=-0.04212+1.286\times E(V-I) -0.09494\times [E(V-I)]^{2}
\end{equation}
\citep{2018A&A...616A...4E}. The extinctions in the Gaia $G$ band were approximately estimated by assuming $A_{G}\backsim A_{V}$ \citep{2018ApJ...866..121H}
and the absolute magnitudes were calculated by using $M_{G}=G+5+5\log(1/\bar{\omega})-A_{G}$, where $\bar{\omega}$ denotes the parallax.

\begin{figure*}
\epsscale{0.9}
\plotone{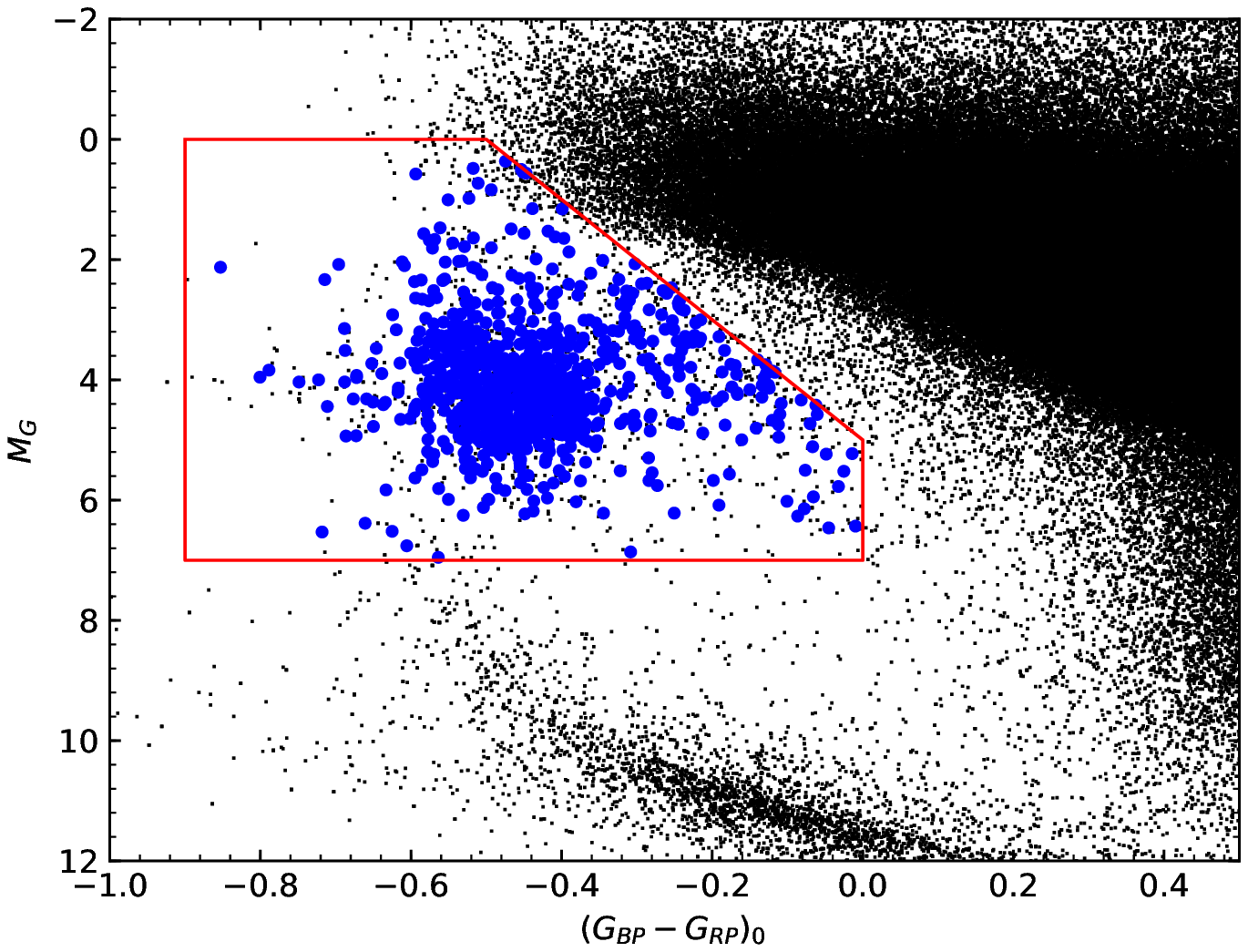}
\caption{Gaia Hertzsprung-Russell diagram of stars selected in GAIA DR2 and LAMOST DR5. The red trapezoid represents selection criteria of hot subdwarf stars and the blue dots denote the positions of the 924 hot subdwarf stars being suitable for a spectral analysis.\label{fig:fig1}}
\end{figure*}

Fig$.\,$\ref{fig:fig1} shows the de-reddened HR diagram.
The figure clearly shows that hot subdwarf stars occupy the region of $-1.0\le (G_{\rm BP}-G_{\rm RP})_{0} \le 0.0$ and $0 \le M_{G}\le7.0$.
To avoid a significant contamination with O and B type MS stars in red end, we defined a trapezoidal selection cut, which
resulted in 1819 hot subdwarf candidates with LAMOST DR5 and Gaia DR2 data. After filtering out bad spectra, MS stars and WDs,
we obtained the resulting 924 stars having suitable spectra (signal-to-noise ratios $S/N \,>\,10$ in magnitude g) for a spectral analysis by visual comparisons to reference spectra of hot subdwarf stars. There are 32 stars in Table$\,$\ref{tab:tab1} that show strong double-lined composite spectra.
Their spectra have noticeable MgI triplet lines at $5170\,$\AA\ or CaII triplet lines at $8650\,$\AA, which were taken as indications of a late type
companion \citep{2009ARA&A..47..211H}. Unfortunately CaII triplet lines are seriously polluted by sky emission lines in some LAMOST spectra.

\begin{figure*}
\epsscale{0.9}
\plotone{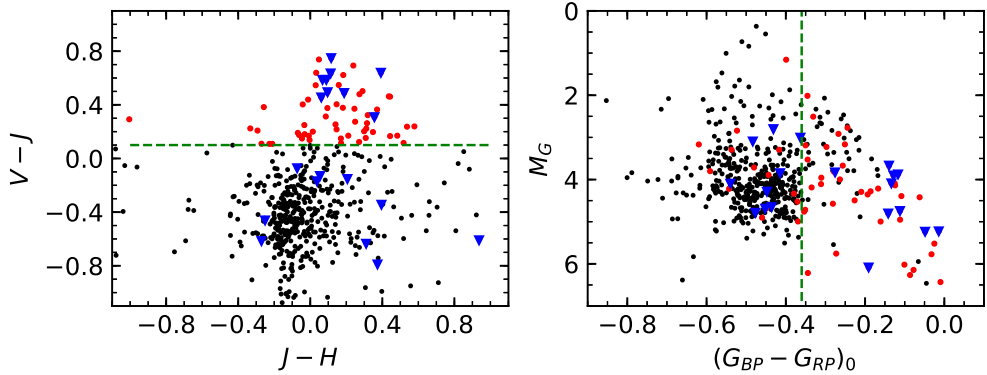}
\caption{Two colour diagram (left panel) and color magnitude diagram (right panel) for 510 hot subdwarf stars having $V-J$ and $J-H$ colors. The blue triangles denote spectra with noticeable MgI triplet lines and the red dots represent spectra with IR excess. The dashed line in the left panel denotes $V-J=0.1$ while that in right panel represents $(G_{\rm BP}-G_{\rm RP})_{0}=-0.36$.\label{fig:fig2}}
\end{figure*}

Due to the radial velocity variations in binary systems, the observed radial velocity measurement at a single epoch cannot give a sufficient
representation of the systems' movement. When we calculated the Galactic space velocities and orbits, we excluded binary systems from our
sample as much as possible. To do this, we adopted two-color diagram using the $V-J$ versus $J-H$ magnitudes and the Gaia HR diagram as demonstrated in Fig$.\,$\ref{fig:fig2}.
Optical $V$ magnitudes were acquired from the Guide Star Catalog (GSC2.3.2; \citealt{2008AJ....136..735L}) and infrared (IR) $J$ and $H$ magnitudes from the Two Micron All-Sky Survey (2MASS; \citealt{2003yCat.2246....0C}).
The left panel of Fig$.\,$\ref{fig:fig2} exhibits a two-color diagram of $V-J$ versus $J-H$ for only 510 hot subdwarf stars in our sample because these
two colors are not available for the other 414 stars. The right panel shows their positions in the Gaia HR diagram.
Following the method of \cite{2017A&A...600A..50G}, we used the color criterium $V-J>0.1$ to distinguish binary systems.
There are 60 binary systems distinguished by $V-J>0.1$ in the $V-J$ versus $J-H$ two-color diagram. The distribution of these 60 binary systems
in the HR diagram of $M_{G}$ versus $(G_{\rm BP}-G_{\rm RP})_{0}$ shows that less than $3\%$ of the composite spectrum binaries are in the range of $(G_{BP}-G_{RP})_{0}<-0.36$
while in other regions (e.g.: $(G_{BP}-G_{RP})_{0}\ge-0.36$) this fraction is more than $50\%$.  After filtering out stars with double-lined composite spectra,
we obtained 747 hot subdwarf stars with radial velocities reliable for calculating their Galactic space motions and orbits in the region of $(G_{BP}-G_{RP})_{0}<-0.36$.

\begin{figure*}
\epsscale{0.9}
\plotone{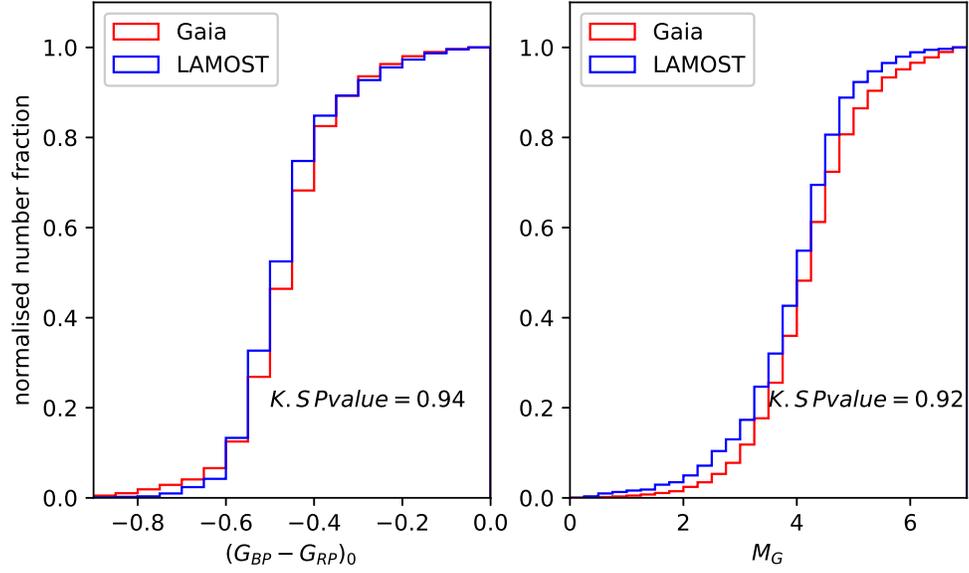}
\caption{Comparisons of the cumulative distribution functions of the Gaia color $(G_{BP}-G_{RP})_{0}$ and absolute magnitudes $M_{G}$ between this study and  \cite{2019A&A...621A..38G}. \label{fig:fig3}}
\end{figure*}

With a global census for the whole sky, homogeneous astrometry and photometry of unprecedented accuracy, Gaia DR2 provides the most complete HR diagram
\citep{2018A&A...616A..10G}, which is a very important tool to check the completeness of the sample.
\cite{2019A&A...621A..38G} presented a catalogue of $39\,800$ hot subdwarf candidates selected from Gaia DR2, which is
the most complete catalogue of hot subdwarf candidates. There are $9\,958$ stars located in our selection area.
Figure$\,$\ref{fig:fig3} shows a comparison of the cumulative functions of the Gaia color $(G_{BP}-G_{RP})_{0}$ and absolute magnitudes $M_{G}$ between
this study and \cite{2019A&A...621A..38G}. Overall, the two sets of the cumulative functions are quite similar and
the Kolmogorov-Smirnov ($K.S.$) test gives a $P$ value of $0.94$ and $0.92$, respectively.

\section{Atmospheric parameters and physical properties} \label{sec:atmo}
\subsection{Model atmosphere calculations}
All the non-LTE H-He composition model atmospheres of hot subdwraf stars were computed with the code {\sc Tlusty} \citep{1995ApJ...439..875H}, version 205.
The corresponding spectral synthesis was performed using the spectrum synthesis code {\sc Synspec} \citep{2011ascl.soft09022H}, version 51.
A detailed description of these programs can be found in the corresponding User's Guides \citep{2017arXiv170601935H,2017arXiv170601937H,2017arXiv170601859H}.
Model atmospheres were calculated in plane-parallel geometry. The horizontally homogeneous atmospheres are in hydrostatic and radiative
equilibrium. The model atoms of H and He were taken from the BSTAR database \citep{2007ApJS..169...83L}. The H line profiles are computed using the Stark
broadening tables of \cite{2009ApJ...696.1755T}. Detailed line profiles of four HeI lines ($\lambda4026$,$\lambda4388$,$\lambda4471$, and $\lambda4922$\AA) are treated using the
special line broadening tables of \cite{1974ApJ...190..315M}, and HeII line profiles are given through the Stark broadening tables of \cite{1989A&AS...78...51S}.

\subsection{Radial velocity}
The published radial velocities of LAMOST spectra are not reliable for hot subdwarf stars, because these stars are not included in LAMOST stellar
templates. Therefore we measured radial velocities with respect to the synthetic spectra of hot subdwarf stars. The observed spectra are cross-correlated with synthetic spectra
templates to determine the radial velocities. All spectra are continuum normalized, Fourier-transformed and convolved with the transform of each template.
For each template, the radial velocity and error is determined by using a Gaussian function to fit the cross-correlation function. The final radial velocity
chosen is the one with the highest confidence value.

\subsection{Spectral analysis} \label{subsec:spec}
Following the method of \cite{2012MNRAS.427.2180N}, atmospheric parameters (effective temperature $T_{\rm eff}$, surface gravity $\log\,g$  and He
abundances $y=n{\rm He}/n{\rm H}$) were measured by fitting the observations with a grid of synthetic spectra. The synthetic spectra were convolved with a Gaussian profile to
degrade the resolution to the LAMOST resolution of $R=1800$. Then they were normalized in $80$ \AA\ sections and fitted to the LAMOST flux calibrated observations. We used the wavelength range
of $3800-7200\,$\AA, which covers all significant H and He lines in LAMOST spectra. We applied the Levenberg-Marquardt minimization algorithm from the \textit{LMFIT}
\footnote{\url{https://lmfit.github.io/lmfit-py/intro.html}} package to determine the best fitting parameters and estimate the standard errors of the free parameters.

\begin{figure*}
\epsscale{1.1}
\plotone{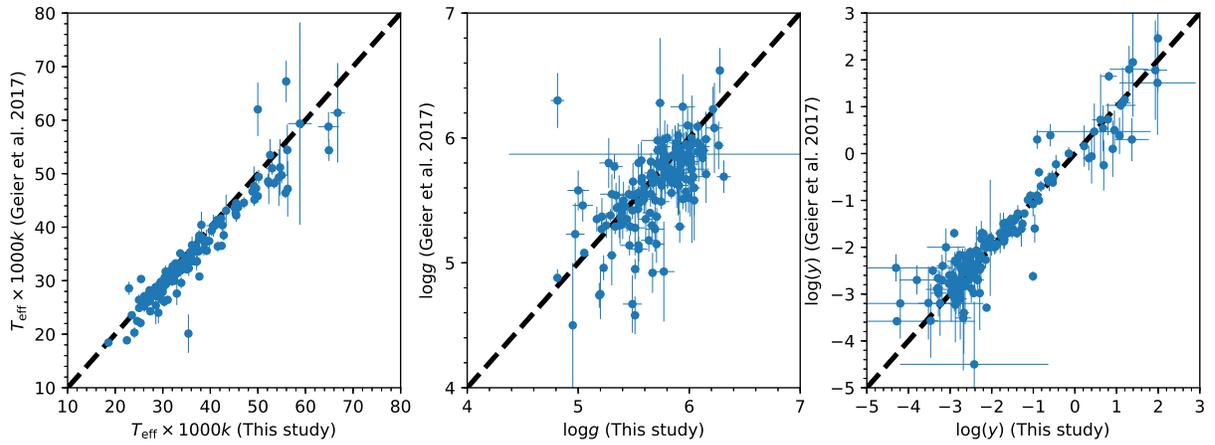}
\caption{Comparisons of atmospheric parameters between this study and the catalog of \cite{2017A&A...600A..50G}.\label{fig:fig4}}
\end{figure*}

A total of 892 single-lined hot subdwarf spectra were analyzed and the rest were left for a forthcoming work.
Table$\,$\ref{tab:tab2} lists the atmospheric parameters of these 892 stars. Figure$\,$\ref{fig:fig4} shows the comparisons
of atmospheric parameters of 163 common stars between this study and \cite{2017A&A...600A..50G}, for which $T_{\rm eff}$, $\log g$ and $\log(y)$ are
simultaneously available. As reported by \cite{2018ApJ...868...70L}, $T_{\rm eff}$ and $\log(y)$ for both sample are quite similar. Although $\log g$
exhibits a larger dispersion than the former two parameters, our results agree with the values presented in the catalogue of \cite{2017A&A...600A..50G}.

\subsection{Physical properties} \label{subsec:phy}
\begin{figure*}
\epsscale{0.9}
\plotone{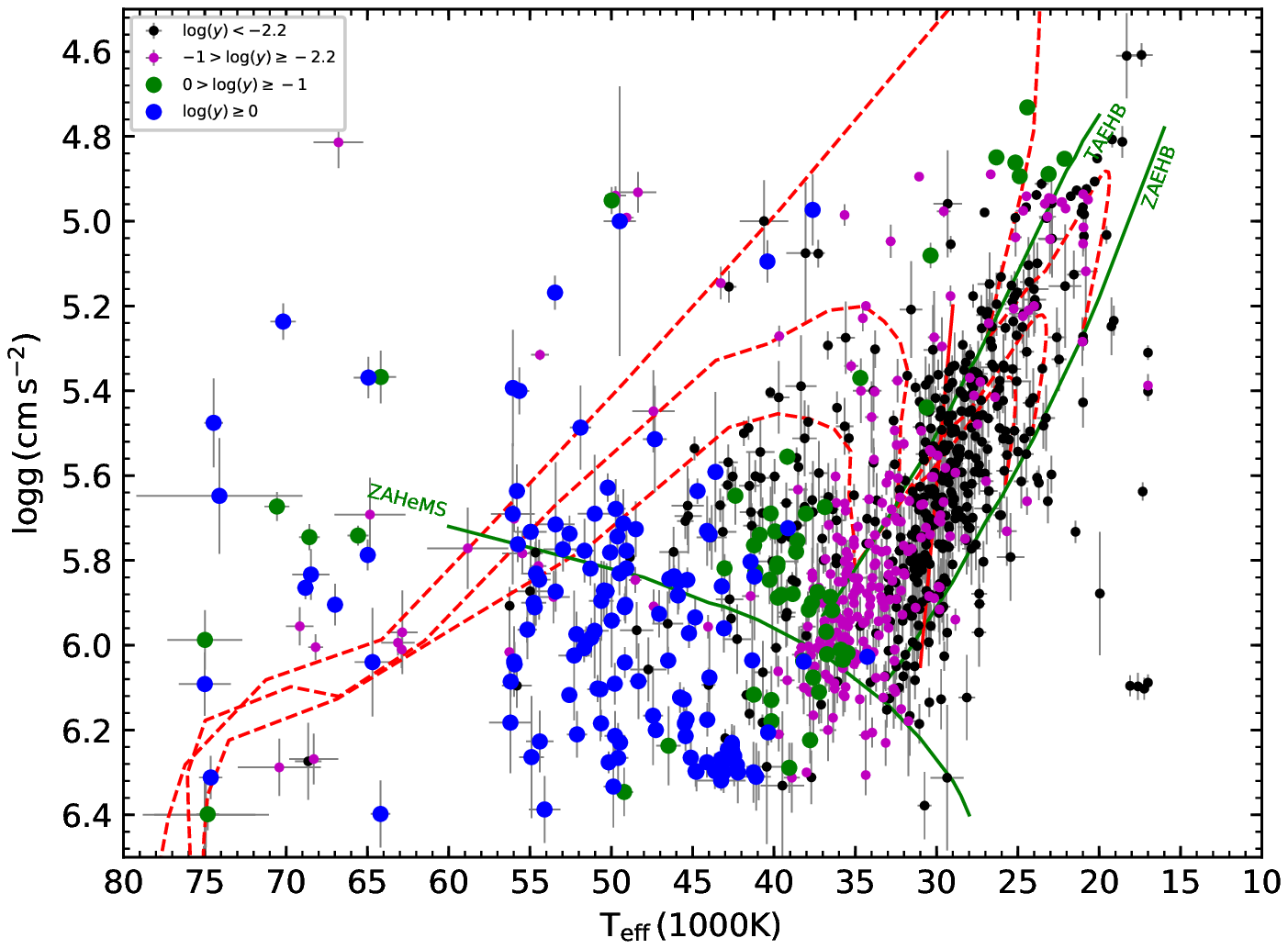}
\caption{$T_{\rm eff}-\log\,g $ diagram. The zero-age EHB (ZAEHB), terminal-age EHB (TAEHB) \citep{1993ApJ...419..596D}, and zero-age He main sequence (ZAHeMS) \citep{1971AcA....21....1P} are marked with the green lines, respectively. The red dashed lines express the evolutionary tracks of \cite{1993ApJ...419..596D} for solar metallicity and subdwarf masses from top to bottom: 0.480, 0.473 and $0.471M_{\odot}$.
The red solid line denotes the observed boundary of slow (to the right) and rapid (to the left) pulsating sdB stars \citep{2010A&A...516L...6C}.\label{fig:fig5}}
\end{figure*}

Figure$\,$\ref{fig:fig5} plots the distribution of hot subdwarf stars in $T_{\rm eff}-\log\,g$ diagram.
We marked the location of the zero age He main sequence (ZAHeMS) given by \cite{1971AcA....21....1P} and the EHB band shown in Fig$.\,5$
in \cite{2012MNRAS.427.2180N}. The EHB band is defined as the region between the zero age extended horizontal branch (ZAEHB) and the terminal age extended horizontal branch (TAEHB) derived from evolutionary tracks of \cite{1993ApJ...419..596D} for solar metallicity and a subdwarf core mass of $0.47M_{\odot}$.
Three evolutionary tracks of \cite{1993ApJ...419..596D} for solar metallicity and
subdwarf masses of 0.471, 0.473, $0.480M_{\odot}$ and the observed boundary of $g-$mode and $p-$mode pulsating sdB
stars \citep{2010A&A...516L...6C} are also marked in Fig$.\,$\ref{fig:fig5}.

Classically, hot subdwarf stars were classified as He-rich and He-deficient with respect to the solar helium abundance $\log{y}=-1$.
Moreover, He-rich and He-deficient stars can also be independently divided into two groups via $\log(y)=0$ and $\log(y)=-2.2$.
As described by \cite{2012MNRAS.427.2180N, 2016ApJ...818..202L}, the two groups of He-deficient stars correspond to $g-$mode
and $p-$mode pulsating sdB stars in EHB in $T_{\rm eff}-\log(g)$ diagram. However, in spite of this correlation the pulsational instability of individual stars cannot be inferred from their position in the $T_{\rm eff}-\log{g}-\log(y)$ parameter space.
Likewise, He-rich stars were classified to extreme He-rich and intermediate He-rich groups via $n_{\rm He}=80\%$ by \cite{2017MNRAS.467...68M}.
Our classification scheme inherently associates the two groups with different formation channels in the $T_{\rm eff}-\log(y)$ diagram.

Figure$\,$\ref{fig:fig5} reveals four different groups of hot subdwarf stars in the $T_{\rm eff}-\log\,g$ diagram via our helium abundance classification scheme.
As seen in both \cite{2012MNRAS.427.2180N} and \cite{2016ApJ...818..202L}, He-deficient sdB stars ($\log(y)<-1$) show two groups on the EHB band.
One is composed of the lower temperature, gravity and He abundance sdB stars ($\log(y)<-2.2$).
They lie to the right of the observed boundary of $g-$mode and $p-$mode pulsating sdB stars and are identified as potential g$-$mode pulsators.
Another group is the higher temperature, gravity and He abundance sdB stars ($-2.2\le\log(y)<-1$) which are on average ten times more He
abundant than the former group. These stars are located to the left of the observed boundary of $g-$mode and $p-$mode pulsating sdB stars and are considered as
potential p$-$mode pulsators. To the left of the EHB band, He-rich stars also exhibit two groups at higher He abundances.
One is consisted of the He-rich sdO stars ($\log(y)\ge 0$) between $40\,000\,$K and $57\,000\,$K near $\log g=6.0$ around the theoretical ZAHeMS.
Compared to LAMOST DR1 \citep{2016ApJ...818..202L}, the five times larger hot subdwarf sample in our work shows clearly that half of the He-rich sdO stars are located below the theoretical ZAHeMS.
Another group is the mix of the intermediate-He sdO/sdB stars ($-1\le\log(y)<0$) around $T_{\rm eff}=38000$ and $\log\,g=5.9$.
In addition, our sample shows a gap between $T_{\rm eff}>57\,000$ and $T_{\rm eff}<63\,000$ where only a few sdO stars can be found and was not reported in LAMOST DR1 \citep{2016ApJ...818..202L}. These observations are consistent with the results of previous reports
\citep{2009PhDT.......273H, 2012MNRAS.427.2180N, 2016ApJ...818..202L, 2018ApJ...868...70L}.

\begin{figure*}
\epsscale{0.9}
\plotone{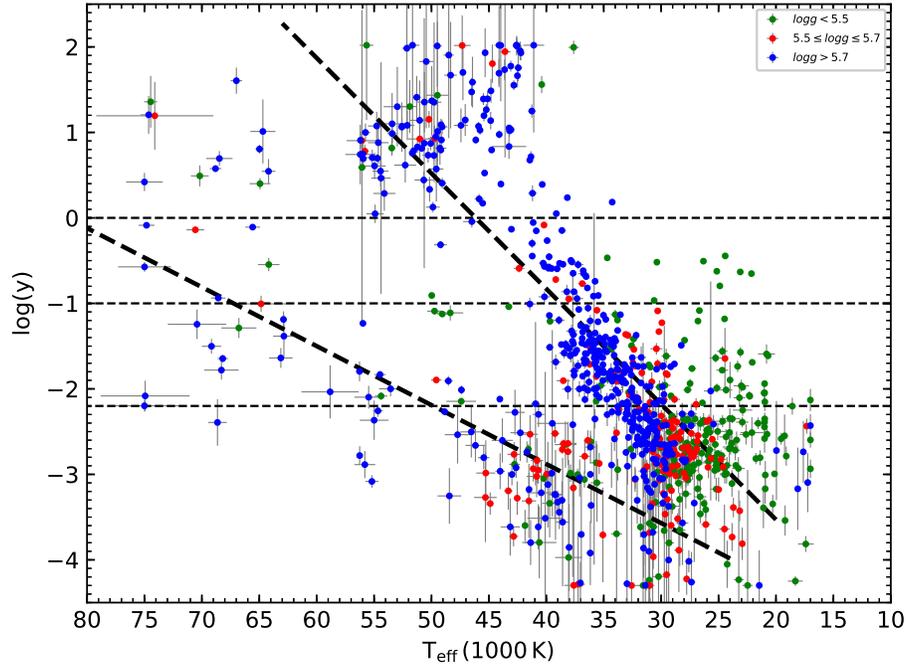}
\caption{Helium abundance versus effective temperature. The thick dashed lines represent the best fitting trends, the top one from \cite{2003A&A...400..939E} and the bottom one from \cite{2012MNRAS.427.2180N}. Three thin dashed lines denote $\log(y)=0$, $\log(y)=-1$, and $\log(y)=-2.2$. \label{fig:fig6}}
\end{figure*}

The $T_{\rm eff}-\log(y)$ diagram is another important parameter space for understanding the formation and evolution of hot subdwarf stars.
Fig$.\,$\ref{fig:fig6} shows the distribution of our sample in the $T_{\rm eff}-\log(y)$ diagram.
Our sample shows the same He sequence as reported by earlier studies \citep{2003A&A...400..939E, 2009PhDT.......273H, 2012MNRAS.427.2180N, 2016ApJ...818..202L,2018ApJ...868...70L}, with
a clear trend having higher He abundances on average at higher temperatures.
In order to compare with previous results, we plot two best-fit trends which can be described by the following relationships:
\begin{equation}
\rm{I:}
\log(y)=-3.53+1.35(T_{\rm{eff}}/10^{4}\rm{K}-2.00),
\end{equation}
\begin{equation}
\rm{II:}
 \log(y)=-4.26+0.69(T_{\rm{eff}}/10^{4}\rm{K}-2.00).
\end{equation}
The former is taken from \cite{2003A&A...400..939E} and the latter from \cite{2012MNRAS.427.2180N}. The majority of hot subdwarf stars
lie near or above the first sequence. Four groups corresponding to those in $T_{\rm eff}-\log\,g$ diagram also appear in this sequence.
The first best-fitting trend is able to match He-deficient stars located in the first sequence, but cannot be suitable for He-rich stars.
Compared with other stars in the first sequence, He-rich stars show a larger scatter in temperature.
Although it seems that some He-rich stars distribute around the first best-fit trend, He-rich stars with $\log(y)\ge0$ follow a contrary tendency: Helium abundance decreases with temperature and approach $\log(y)=-0.5$ and He-rich stars with $-1\le\log(y)<0$ also deviate from the best-fit trend. This distribution suggests that these two groups of He-rich stars may origin from different formation channels.
These results are in good agreement with previous studies \citep{2007A&A...462..269S, 2009PhDT.......273H, 2012MNRAS.427.2180N, 2016ApJ...818..202L,2018ApJ...868...70L}. He-deficient sdB/O stars form the second sequence. Although the distribution of stars in the second sequence shows a larger scatter, they can be matched with the second best-fitting trend. In addition, as reported by \cite{2012MNRAS.427.2180N} and \cite{2016ApJ...818..202L}, over $T_{\rm eff}=40\,000\,$K, our sample shows a gap between $\log(y)>-1.5$ and $\log(y)<0$. There are only a few stars with relatively low $\log\,g$ in this gap.

Although many scenarios have been put forward to explain the formation and evolution of hot subdwarf stars, none of them appear entirely satisfactory.
\cite{2002MNRAS.336..449H, 2003MNRAS.341..669H} proposed three main formation channels: the common envelope (CE) ejection, the stable Roche lobe overflow (RLOF),
and the merger of double helium white dwarf (WDs) binaries and predicted distinct properties of hot subdwarf stars from these channels.
Due to the difference in He abundance, we can outline two groups of sdB stars in the $T_{\rm eff}-\log(g)$ diagram and they also correspond to potential $g-$mode and $p-$mode sdB pulsators, respectively. Based on the analysis of long-period composite
spectrum binaries (sdB+F/G) \citep{2012MNRAS.427.2180N}, we associated the lower temperature and surface gravity group with the CE ejection
channels and the high temperature and gravity group with the Roche-lobe overflow channel in the LAMOST DR1 sample \citep{2016ApJ...818..202L}.
However, more recent observations \citep{2015MNRAS.450.3514K, 2015A&A...576A..44K} found that both short-and long period binaries occur in each group,
implying that these groups have a mixture of stars with different formation history. Therefore, further investigations are needed to verify whether one can infer
to the yields of various formation channels from these two groups.

As described in \cite{2017MNRAS.467...68M}, He-rich stars with $-1\le\log(y)<0$ present more of a challenge. In the $T_{\rm eff}-\log\,g$ and $T_{\rm eff}-\log(y)$
diagrams, He-rich stars with $-1\le\log(y)<0$ show a group as marked in Figure 8 in \cite{2016ApJ...818..202L}.
In this region \cite{2016ApJ...818..202L} identified only seven stars from LAMOST DR1 while we obtained 54 stars from LAMOST DR5.
The correlations between He abundance and temperature for these stars are different from He-rich stars with $\log(y)\ge0$
and from He-deficient stars. There are two possible and equally reasonable formation channels for these stars: the hot flasher scenario with tidally enhanced stellar
wind in binary evolution \citep{2015MNRAS.449.2741L} and the mergers of helium-core white dwarfs with low-mass MS stars \citep{2017ApJ...835..242Z}.
Although theoretical evolutionary tracks of these two formation channels can cover these stars in both the $T_{\rm eff}-\log\,g$ and $T_{\rm eff}-\log(y)$ diagrams,
current observations cannot provide enough evidence to distinguish them. The distribution of He-rich stars with $\log(y)\ge0$ are similar in other
samples \citep{2009PhDT.......273H, 2012MNRAS.427.2180N,2016ApJ...818..202L,2018ApJ...868...70L}. Binary population synthesis models show that the double
white dwarf merger channel can explain He-rich stars with $\log(y)\ge0$. However, a recent study by \cite{2012MNRAS.419..452Z} shows that the evolutionary tracks
from extensive calculations for the double HeWDs merger can only cover these stars above or near the HeMS in the $T_{\rm eff}-\log\,g$ diagram, but are not suitable
for stars located below the HeMS. More than half of the He-rich stars with $\log(y)\ge0$ in our sample lie below the HeMS in $T_{\rm eff}-\log\,g$ diagram.

As discussed by \cite{2016ApJ...818..202L}, the reasons for the correlations between helium abundance and temperature and the different structure of
the sequences in the $T_{\rm eff}-\log(y)$ panel are not fully understood. The evolutionary sketches of \cite{2012MNRAS.427.2180N}
are able to provide a qualitative picture for the LAMOST DR5 sample, they should be explored in detail with numerical models.

\section{Space distribution} \label{sec:spac}
With the Gaia parallaxes ($\bar{\omega}$) and equatorial coordinates ($\alpha$ and $\delta$), we calculated the position of hot subdwarf stars
in right-handed Cartesian Galactic coordinates in which the $X$-axis is positive towards the Galactic Center, the $Y$-axis in direction of Galactic
rotation and the $Z$-axis towards the North Galactic Pole. The distance of the Sun from the Galactic Center is taken to be 8.4 kpc \citep{2010MNRAS.403.1829S}.
In order to propagate the error on the input parameters, a Monte Carlo simulation was performed. For an input parameter $x_{0}$, with error $\sigma$,
the error of the output parameter $f(x)$ can be obtained as follows: first $1\,000$ random $x-$values, obeying a Gaussian distribution, with mean value $x_{0}$ and standard deviation $\sigma$, are generated. Then $f(x)$ is computed for all these $x$ values and the mean value and error of $f(x)$ are
calculated. We applied such a Monte Carlo simulation to calculate the errors of the Galactic position components.

\begin{figure*}
\epsscale{1.2}
\plotone{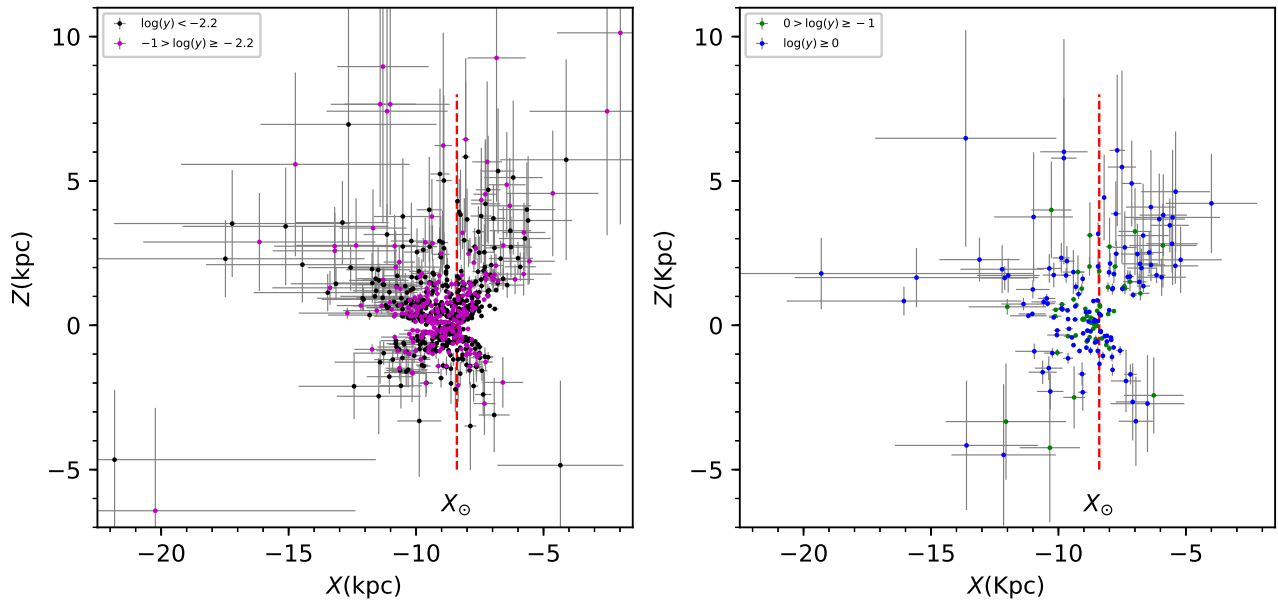}
\caption{The positions of 892 non-composite spectra hot subdwarf stars in Cartesian Galactic $X-Z$ coordinates. He-deficient stars are shown in the left panel and He-rich stars are displayed in the right panel. The dashed line denotes the solar position. \label{fig:fig7}}
\end{figure*}

Figure$\,$\ref{fig:fig7} shows the space positions of the four hot subdwarf helium groups in the $X-Z$ diagrams.
The left panel of Fig$.\,$\ref{fig:fig7} reflects that the space distributions of the two groups of He-deficient hot subdwarf
stars do not show any obvious differences. The vast majority of He-deficient stars assembles near the disk and only a few stars disperse in the halo.
The star density of the disk is much higher than of the halo and exhibits a sharp cut-off at $|Z|\sim1.5\,{\rm kpc}$, which is considered as the vertical
scale height for the thick disk \citep{2017MNRAS.467.2430M}. The space distribution of He-rich stars are shown in the right panel of Fig$.\,$\ref{fig:fig7}.
The space density of the two group of He-rich stars shows a larger dispersion than the groups of He-deficient stars from the thin disk to halo.
Moreover, these two groups of He-rich stars also exhibit a clear difference in space distribution. At $|Z|>1.5\,{\rm kpc}$, the star density
of He-rich stars with $\log(y)\ge0$ is much higher than that of stars with $-1\le\log(y)<0$. Their space distribution also indicates that the two groups
of He-rich stars are likely to origin from different formation channels. In addition, there exists a sharp cut-off in the star density at $|Z|\sim1.5\,{\rm kpc}$
for these four helium groups and stars with $|Z|>1.5\,{\rm kpc}$ show a diverse range in space distributions, which suggests that hot subdwarf stars
with $|Z|>1.5\,{\rm kpc}$ have different kinematic origins.

\section{Kinematics} \label{sec:kine}
As noted in subsection$\,$\ref{subsec:samp}, the single measurement of the radial velocity cannot represent the systemic radial velocity of a hot
subdwarf binary. In order to filter out binary systems from our sample, we selected 747 hot subdwarfs in the region of
$(G_{BP}-G_{RP})_{0}<-0.36$ as targets for kinematics, because the percentage of hot subdwarf stars with a companion in the selected region
is estimated to be less than $3\%$ via the $V-J$ versus $J-H$ diagram.

\subsection{Calculation of Galactic velocities} \label{subsec:cal}
Combining the five-parameter astrometric solutions ($\alpha$,$\delta$, $\bar{\omega}$, $\mu_{\alpha}\cos\delta$, and $\mu_{\delta}$) provided by Gaia DR2
with the radial velocities ($RV$) measured from LAMOST spectra, we computed space velocities of hot subdwarf stars in the Cartesian Galactic coordinates with
the \textit{Astropy} Python package. We took a right-handed Cartesian Galactic coordinate in which the $X$-axis is positive towards the Galactic Center.
For the Cartesian Galactic velocities, we defined three components $U$, $V$, and $W$, where $U$ is the Galactic radial velocity positive towards the Galactic
centre, $V$ is the Galactic rotational velocity in the direction of the Galactic rotation, and $W$ is the component oriented towards the North Galactic Pole.
The distance of the Sun from the Galactic centre is set to be 8.4 kpc and its velocity with respect to the local standard of rest (LSR) is taken to be
($U_{\odot}$,$V_{\odot}$,$W_{\odot}$)$=$(11.1, 12.24, 7.25)$\,{\rm km\,s^{-1}}$ \citep{2010MNRAS.403.1829S}. The velocity of the LSR is set to be $242\,{\rm km\,s^{-1}}$
\citep{2013A&A...549A.137I}. The errors of kinematic parameters are computed with the Monte Carlo simulation described in Sec$.\,$\ref{sec:spac}.

\subsection{Galactic velocities and velocity dispersion} \label{subsec:gal}
\begin{figure*}
\epsscale{0.9}
\plotone{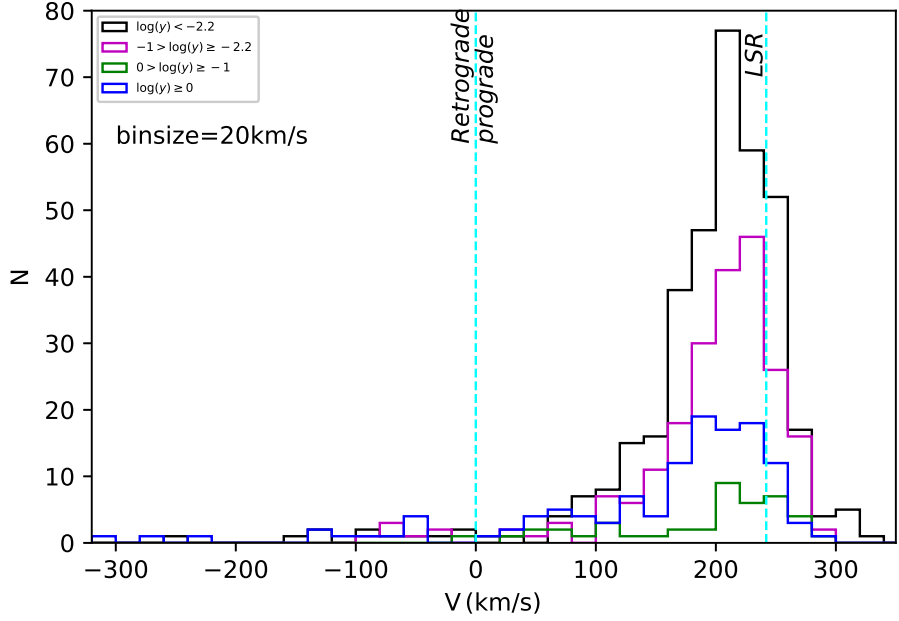}
\caption{Histogram of the Galactic rotational velocities of 747 hot subdwarf stars selected from Gaia DR2 and LAMOST DR5. The dashed left line at a Galactic rotational velocity of zero is to highlight stars with retrograde motion and the dashed right line at a Galactic rotational velocity of $242{\rm km\,s^{-1}}$ represents the LSR.
\label{fig:fig8}}
\end{figure*}

Figure$\,$\ref{fig:fig8} shows the histogram of the rotational velocity, $V$, for 747 hot subdwarf stars. As described by \cite{2017MNRAS.467...68M},
these stars display a peak at $220\,{\rm km\,s^{-1}}$ near the LSR and a tail extending into negative rotational velocities.
These features have been seen in the sample of 114 sdB stars analysed by \cite{2004A&A...414..181A} as well as in the sample of 398
DA WDs from the SPY project \citep{2006A&A...447..173P}. These indicate that the rotational velocities of the vast majority of
these stars are similar to disk stars.

\begin{figure*}
\epsscale{0.9}
\plotone{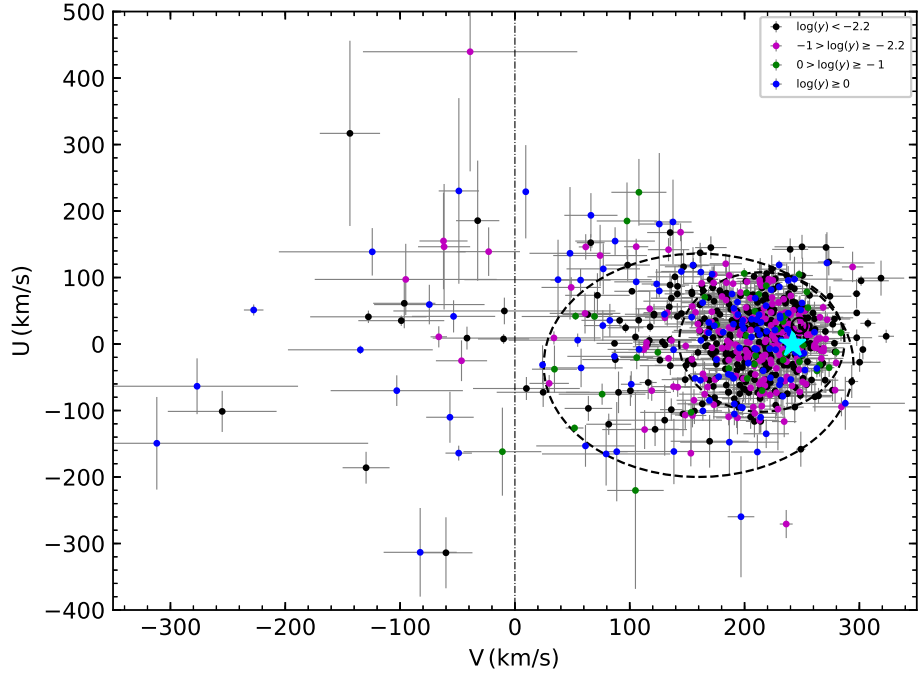}
\caption{$U-V$ velocity diagram for 747 hot subdwarf stars selected in Gaia DR2 and LAMOST DR5. Two dashed ellipses represent the $3\sigma$ limits for the thin disk and thick disk populations \citep{2006A&A...447..173P}. The cyan star denotes the Local Standard of Rest (LSR).
\label{fig:fig9}}
\end{figure*}

The $U-V$ velocity diagram is a classical tool for kinematic investigations. The distribution of our sample in the $U-V$ velocity diagram is shown in
Fig$.\,$\ref{fig:fig9}. As shown in Figure$\,$1 in \cite{2017MNRAS.467...68M}, two dotted ellipses denote the $3\sigma-$limits of thin and thick disk
WDs \citep{2006A&A...447..173P}, respectively. There are about $72\%$ of our stars located in the $3\sigma-$limit of thin disk stars, $15\%$ lie in
the region between the $3\sigma-$limits of thin and thick disk stars and $13\%$ are located outside of the $3\sigma-$limit of thick disk stars.
He-deficient stars tend to group around the LSR, while He-rich stars are wildly scattered in the whole region.

\begin{figure*}
\epsscale{0.85}
\plotone{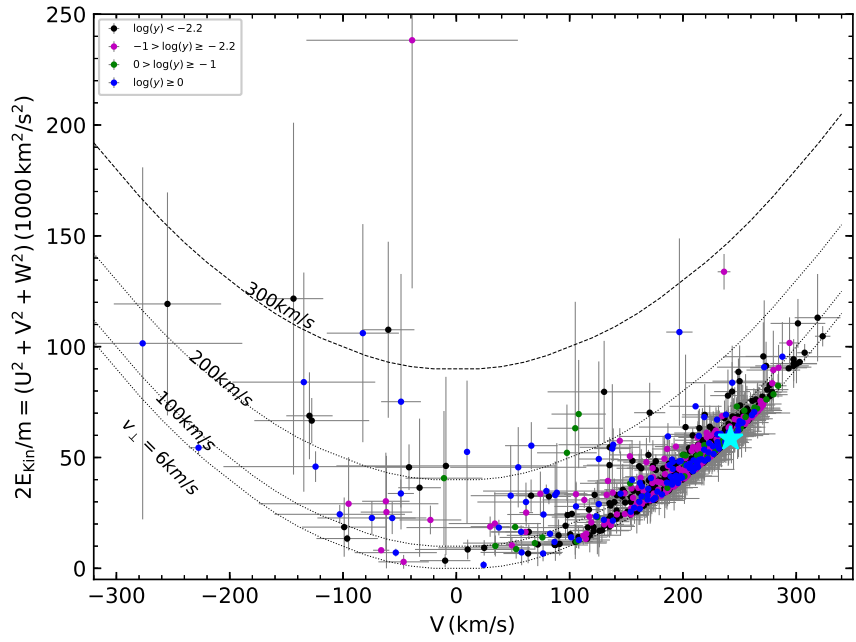}
\caption{Galactic rotational velocity against the total kinetic energy for 747 hot subdwarf stars selected in Gaia DR2 and LAMOST DR5. The parabolic curves denote lines of equal velocity values $V_{\perp}=(U^{2}+V^{2})^{1/2}$. The cyan star denotes the Local Standard of Rest (LSR).
\label{fig:fig10}}
\end{figure*}

The kinetic energy $2E_{\rm kin}/m=U^{2}+V^{2}+W^{2}$ versus rotational velocity ($V$) diagram, is another important tool to analyse the kinematics
 of these stars, and displayed in Fig$.\,$\ref{fig:fig10}. The higher the value of the kinetic energy $2E_{\rm kin}/m$, the more elliptic is the orbit of the star. The parabolic curves mark the isovelocity curve $V_{\perp}=(U^{2}+V^{2})^{1/2}$ in Fig$.\,$\ref{fig:fig10}. For low values of $V_{\perp}$,
 the deviation from LSR provides information about the kinetic temperature. As shown in \cite{2017MNRAS.467...68M}, most of stars group around the LSR
 on a "banana" shaped region alongside the $V_{\perp}=0{\rm \,km\,s^{-1}}$ isovelocity curve, which means that they are kinematically cool. Another interpretation
 for this clustering is that most stars are near their orbital turning points \citep{2004A&A...414..181A}. One can also see that a few stars are located
 further away from the $V_{\perp}=0 {\rm \,km\,s^{-1}}$ contour, in which He-rich stars with $\log(y)\ge0$ have a higher proportion. These stars are kinematically
 hot and likely to have a more eccentric orbit. Moreover, the sample also shows a sharp cut near $110{\rm \,km\,s^{-1}}$. Stars located at the left of this velocity are
 more sparser than those at the right of this $V_{\perp}$ and they belong to the halo population \citep{2004A&A...414..181A}. In this region, He-rich
 stars with $\log(y)\ge0$ have also very high proportion.

The mean values and standard deviations of the galactic velocities for the four hot subdwarf helium groups are listed in Table \ref{tab:tab4}.
He-rich stars with $\log(y)\ge0$ exhibit the largest standard deviation of the galactic velocity components and He-rich stars with $-1\le\log(y)<0$ display
the second largest standard deviation. The two groups of He-deficient stars show a similar values of standard deviation.
To conclude, He-rich hot subdwarf stars with $\log(y)\ge0$ show a diverse range of kinematic velocities, which indicates that these stars are likely from different formation channels.
The triples with a hot subdwarf star component had been discovered \citep{2002A&A...383..938H,2012ApJ...758...58B,2015A&A...576A..44K}, but they may only make a very small contribution to the kinematic diversity. \cite{2010ApJS..190....1R} showed that nearly $26\%$ of all solar-type binaries are in hierarchical triples. However, \cite{2015A&A...576A..44K} found a fraction of $2.1\%$ for triples in their hot subdwarf binary sample. To keep long term stability, the hierarchical triples mainly produce the hot subwarf binaries with a third component in a wide orbit \citep{2002MNRAS.336..449H, 2019MNRAS.485.2562H}. Aside from potentially advancing the formation process via the Kozai mechanism, the hierarchical triples have no effects on the formation of hot subwarf stars via the binary evolution channels \citep{2012ApJ...758...58B}. The wide outer orbit may only cause a very small velocity dispersion. The kinematic diversity may be further complicated by dynamical interactions, like stars that were ejected from hierarchical triple systems. However, the probability of these events is very low for field hot subdwarf stars.

\section{Galactic orbits } \label{sec:orb}

\subsection{Calculation of orbits} \label{subsec:orbc}
\begin{figure*}
\epsscale{1.0}
\plottwo{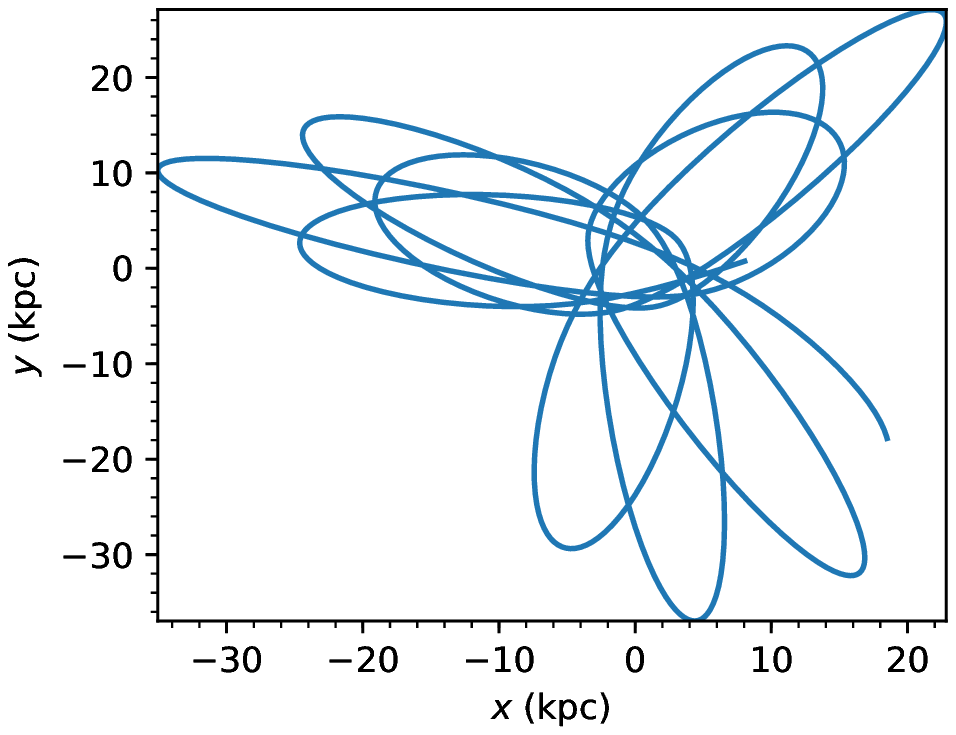}{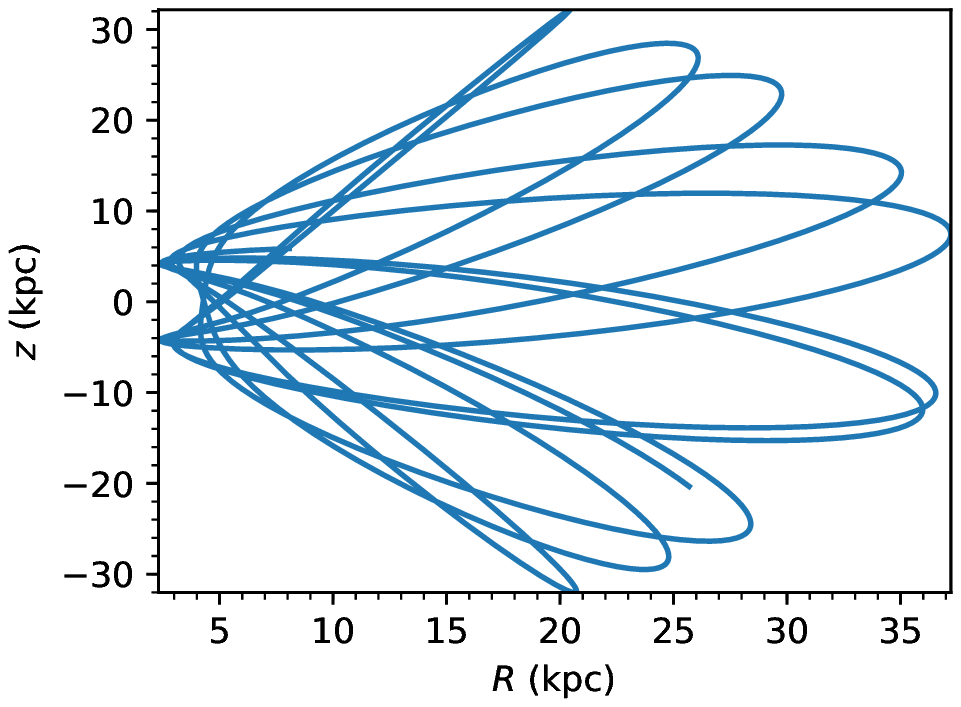}
\caption{Meridional sections of the orbit of LAMOST ${\rm J132300.27+310750.8}$ identified as a halo star.
\label{fig:fig11}}
\end{figure*}

The orbits were calculated using \textit{Galpy}, a Python package for galactic dynamics calculations \citep{2015ApJS..216...29B}.
We took the Milky Way potential "MWpotential2014" for our orbit integrations. The standard MWpotential2014 has a power-law bulge with an exponential cut-off,
an exponential disk and a power-law halo component \citep{2015ApJS..216...29B}. The distance of the Sun from the Galactic Center is
set to be 8.4$\,$kpc and the velocity of the LSR is taken to be $242\,{\rm km\,s^{-1}}$ \citep{2013A&A...549A.137I}.
The time of orbit integration was set to 5 Gyrs in steps of 1 Myr. Fig$.\,$\ref{fig:fig11} shows an example, where
$R=\sqrt{X^{2}+Y^{2}}$ is the radius of the motion in the galactic plane.

From the shape of the orbits we extracted the apocentre ($R_{\rm ap}$), pericentre ($R_{\rm peri}$), eccentricity ($e$), maximum vertical amplitude ($z_{\rm max}$),
normalised z-extent ($z_{n}$) and z-component of the angular momentum ($J_{z}$). These quantities are listed in Table$\,$\ref{tab:tab5}.
$R_{\rm ap}$ and $R_{\rm peri}$ are the maximum and minimum distances from the Galactic centre. We defined the eccentricity by
\begin{equation}
e=\frac{R_{\rm ap}-R_{\rm peri}}{R_{\rm ap}+R_{\rm peri}},
\end{equation}
and the normalised z-extent by
\begin{equation}
z_{n}=\frac{z_{\rm max}}{R(z_{\rm zmax})},
\end{equation}
where $R$ is the galactocentric distance.
The errors of orbital parameters are obtained with the Monte Carlo simulation described in Sec$.\,$\ref{sec:spac}.
For each star $1\,000$ sets of input values with a Gaussian distribution were simultaneously generated and the output parameters together
with their errors were computed.

\subsection{The orbital parameters}\label{subsec:orbpar}
\begin{figure*}
\epsscale{0.9}
\plotone{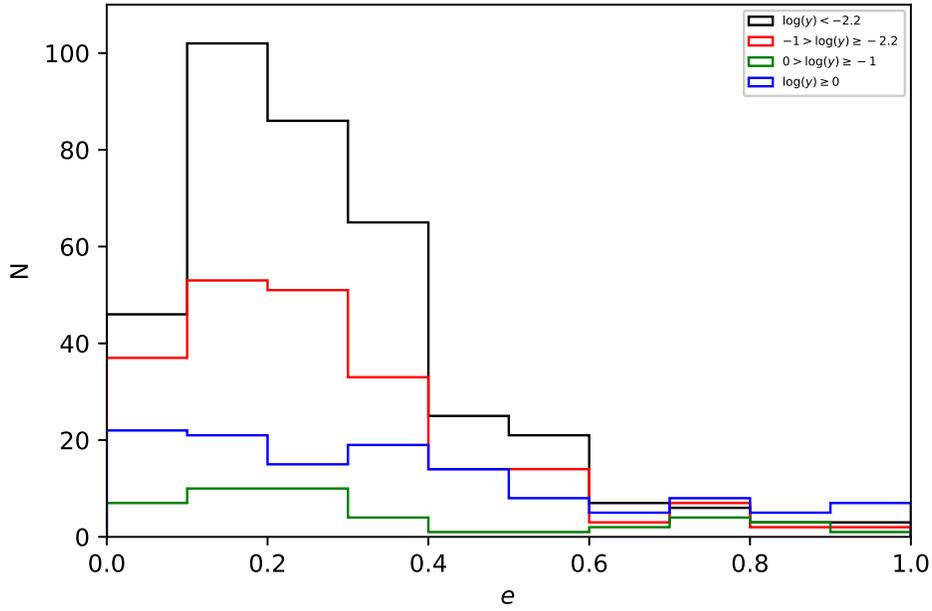}
\caption{Histograms of the eccentricity distribution of 747 hot subdwarf stars selected from Gaia DR2 and LAMOST DR5 .\label{fig:fig12}}
\end{figure*}

Table \ref{tab:tab5} gives the mean values and standard deviations of the orbital parameters: eccentricity, normalised z-extent, maximum vertical amplitude, apocentre and pericentre. He-rich stars with $\log(y)\ge0$ show the largest
standard deviation values for these orbital parameters in all four hot subdwarf helium groups. Figure$\,$\ref{fig:fig12} shows histograms of the eccentricity
distributions of the four hot subdwarf helium groups. He-deficient stars have a peak around $e\approx0.2$, where the majority of disk stars are located.
While He-rich stars display a distribution that is spread over the whole range. These findings are in good agreement with the results of \cite{2017MNRAS.467...68M}.

\begin{figure*}
\epsscale{0.9}
\plotone{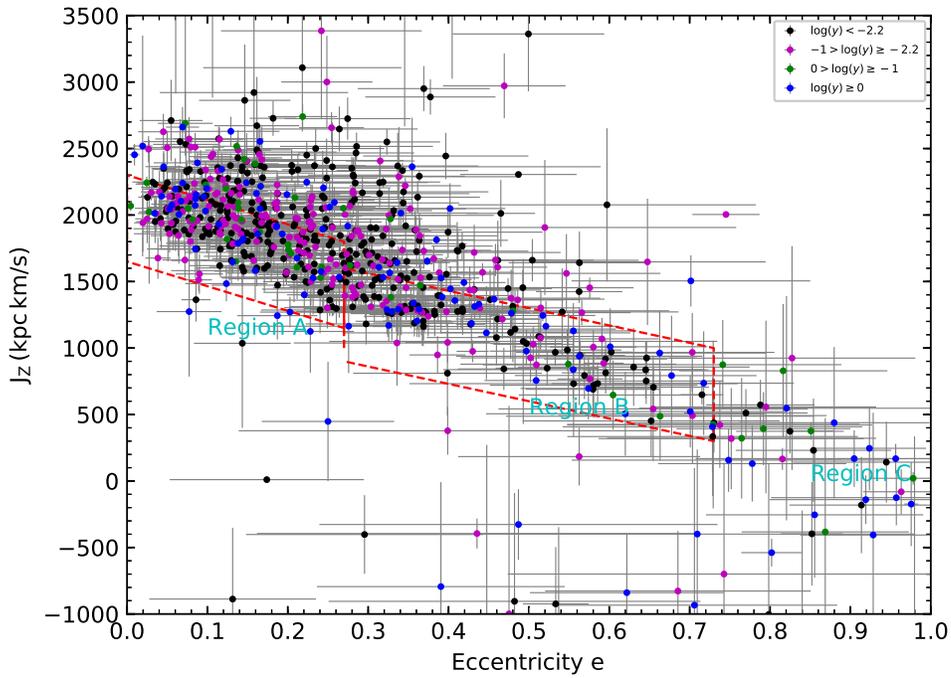}
\caption{Z-component of the angular momentum versus eccentricity ($J_{Z}-e$). Two parallelograms represent Region A (thin disk) and Region B (thick disk) \citep{2006A&A...447..173P}.
\label{fig:fig13}}
\end{figure*}

Both the eccentricity and the z-component of the angular momentum are important orbital parameters to distinguish different populations. The $J_{z}-e$ diagram is shown in Fig$.\,$\ref{fig:fig13}. We also plotted the two regions defined by \cite{2003A&A...400..877P}:
Region A confines thin disk stars clustering in an area of low eccentricity, and $J_{z}$ around $1\,800\,{\rm kpc\,km\,s^{-1}}$, Region B encompasses thick disk stars having
higher eccentricities and lower angular momenta. Outside of these two regions, we defined a region C, in which there are halo star candidates.
Figure\ref{fig:fig13} shows a continuous distribution of stars from Region A to Region B and there is no obvious dichotomy.
Only a few stars lie in Region C and they show a noticeable gap between Region B and Region C.
These are in good agreement with the findings of \cite{2013ApJ...779..115B}.

\section{Galactic Population classifications}\label{sec:pop}
\begin{figure*}
\epsscale{0.8}
\plotone{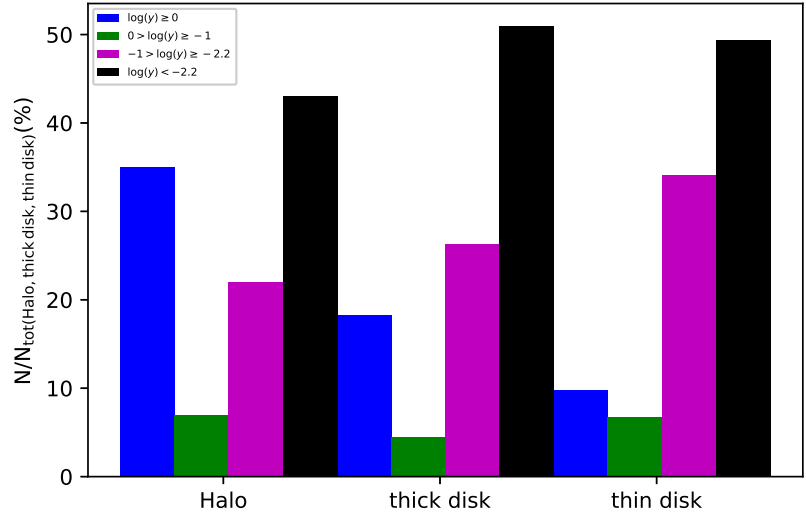}
\caption{The fractional distributions of the four hot subdwarf helium groups defined in Section \ref{subsec:phy} for the halo, thick disk and thin disk populations. \label{fig:fig14}}
\end{figure*}

The $U-V$ diagram, $J_{z}-e$ diagram and the maximum vertical amplitude were used to distinguish the populations of hot subdwarfs.
Following \cite{2017MNRAS.467...68M}, we adopted the classification scheme of \cite{2003A&A...400..877P, 2006A&A...447..173P} and
added the vertical scale height $Z\sim1.5{\rm kpc}$ for thick disk stars as the cut-off height of the thin disk \citep{2017MNRAS.467.2430M}.
All orbits have been visually inspected to supplement the automatic classifications. We applied the following check-list to ensure correct population assignments:
\begin{enumerate}
  \item Stars that belong to the thin disk lie within the $3\sigma$ thin disk contour in the $U-V$ diagram and Region A in the $J_{z}-e$ diagram.
  The extensions of their orbits in the $R$ and the $Z-$directions are small and have $z_{\rm max}<1.5\,{\rm kpc}$.
  \item Stars situated within the $3\sigma$ thick disk contour and in Region B  have been classified as thick disk stars.
  Their orbits are more extended in the $R$ and the $Z-$directions than that of thin disk stars, but they do not cover such a large region as halo stars.
  \item Stars that lie outside Region A and B as well as outside the $3\sigma$ thick disk contour are classified as halo stars. Their orbits show high extensions in $R$ and $Z$. Stars with an extension in $R$ larger than $18\rm{kpc}$, or the vertical distance from the Galactic plane $Z$ larger than $6\rm{kpc}$, are also qualified as halo stars.
\end{enumerate}
Table$\,$\ref{tab:tab6} lists the number of stars in the four hot subdwarf helium groups classified as halo, thin or thick disk stars and
Figure$\,$\ref{fig:fig14} shows their fractions in the halo, thin disk and thick disk.

In fact, the different Galactic populations (thin disk, thick disk, and halo) represent different stellar population ages \citep{2017ApJS..232....2X}.
With binary population synthesis \citep{2002MNRAS.336..449H, 2003MNRAS.341..669H}, \cite{2008A&A...484L..31H} presented the fractions of hot subdwarf
stars from three different formation channels (stable RLOF, CE ejection, and the merger of double HeWDs) at various stellar population ages.
The distribution of stars in the $T_{\rm eff}-\log\,g$ and $T_{\rm eff}-\log(y)$ diagrams argue that He-rich hot subdwarf stars with $\log(y)\ge0$ origin mainly
from the merger channel. Figure \ref{fig:fig14} reflects that the fraction of hot subdwarf stars with $\log(y)\ge0$ increases
from $\sim10\%$ in the thin disk to $\sim33\%$ in the halo, which agrees with the predictions of the merger channel.

For He-deficient stars, we outlined two groups in the $T_{\rm eff}-\log\,g$ and $T_{\rm eff}-\log(y)$ diagrams separated by a gap in He abundance at $\log(y)=-2.2$.
As already discussed in section \ref{subsec:phy}, the formation channels of these two groups are not fully understood in light of the currently available observations.
The fraction of He-deficient stars with $-2.2\le\log(y)<-1$ decrease from $\sim34\%$ in the thin disk to $\sim26\%$ in the halo, which is in good agreement
with the predictions of the CE ejection channel. While the fraction of He-deficient stars with $\log(y)<-2.2$ decrease from $\sim50\%$ in the thin disk
to $\sim40\%$ in the halo. The low fraction of He-deficient stars with $\log(y)<-2.2$ in the thin disk is mainly due to a selection effect in our sample.
All the kinematic sample was selected by setting the criterion of $(G_{BP}-G_{RP})_{0}<-0.36$, which is suitable to decrease the pollution from
binary systems, but also sorts out He-deficient stars at the low temperature side and underestimates their fraction. When the selection effect is taken into account, the fraction of He-deficient stars $\log(y)<-2.2$ agrees with the predictions of the stable RLOF channel.

However, as described by \cite{2017MNRAS.467...68M}, the formation of He-rich hot subdwarf stars with $-1\le\log(y)<0$ is also a puzzle.
The fraction is $\sim7\%$ in the halo after decreasing from $\sim7\%$ in the thin disk to $\sim4\%$ in the thick disk, which implies that He-rich hot subdwarf stars with $-1\le\log(y)<0$ in thin disk and halo may have different formation channels.
Population synthesis results by \cite{2017ApJ...835..242Z} show that the formation of hot subdwarf stars via
the merger channel of HeWD$+$MS occurs at a rate of about $20\%$ that of stars with $\log(y)\ge0$ generated by the HeWD$+$HeWD merger channel.
But this formation channel produces only $40\%$ of He-rich hot subdwarf stars with $-1\le\log(y)<0$.
The relative number ratio of He-rich hot subdwarf stars with $-1\le\log(y)<0$ and $\log(y)\ge0$ monotonously decreases from $69\%$ in the thin disk to $20\%$ in the halo. This relative number ratio in the halo is more than twice the population synthesis results presented by \cite{2017ApJ...835..242Z}, even more than six times in the thin disk, which implies that the formation channel of HeWD$+$MS mergers may not be the main contribution of hot subdwarf stars with $-1\le\log(y)<0$.

\section{Conclusions} \label{sec:conc}
With Gaia DR2 and LAMOST DR5, we have spectroscopically identified 924 hot subdwarf stars, among which 32 stars show strong composite spectra.
We have measured the atmospheric parameters (effective temperature $T_{\rm eff}$, surface gravity $\log\,g$, and He abundances $y=n({\rm He})/n({\rm H})$)
of 892 single-lined stars by simultaneously fitting the profiles of H and He lines in the wavelength range of $3800-7200$\AA\ using synthetic
spectra computed from non-LTE {\sc Tlusty} model atmospheres. The physical properties of the stars have been discussed based on their positions in the $T_{\rm eff}-\log\,g$ and $T_{\rm eff}-\log(y)$ diagrams.
We outlined four hot subdwarf groups in the $T_{\rm eff}-\log\,g$ diagram through our He abundance classifications. Two well defined He-deficient
star groups can be located in the EHB band, at the two sides of the observed boundary between $p-$mode and $g-$mode sdB pulsators. Due to the larger hot subdwarf sample than in
LAMOST DR1, He-rich stars with $\log(y)\ge0$, as well as intermediate He-rich stars with $-1\le\log(y)<0$ show up more clearly. The latter group was
not noticeable in LAMOST DR1 \citep{2016ApJ...818..202L}. Over half of the He-rich stars with $\log(y)\ge0$ are found below the HeMS.
In the $T_{\rm eff}-\log(y)$ diagram, our sample reveals the two known parallel sequences with an overall tendency of having higher helium abundances at higher temperatures.
However, in the first trend He-rich stars with $\log(y)\ge0$ appear to follow an opposite trend and He-rich stars with $-1\le\log(y)<0$ deviate from the best-fitting trend. These results are consistent with previous reports  \citep{2009PhDT.......273H,2012MNRAS.427.2180N, 2016ApJ...818..202L,2018ApJ...868...70L}.

We also presented the space distributions of the 892 single-lined stars in the Milky Way. The vast majority of the stars are found to group within $|Z|\le1.5\,$kpc and shows a sharp cut-off at $|Z|\sim1.5\,$kpc. He-deficient and He-rich
stars exhibit different space distributions. At $|Z|\le1.5\,$kpc, He-deficient stars have a higher space density than He-rich stars.
The two groups of He-deficient stars show similar space distributions, but the two groups of He-rich stars display noticeable differences
in their space distributions. At $|Z|>1.5\,$kpc, He-rich stars with $\log(y)\ge0$ have a higher space density than the ones with $-1\le\log(y)<0$.
These results indicate that the two groups of He-rich stars are from different formation channels.

We have selected 747 hot subdwarfs in the region of $(G_{BP}-G_{RP})_{0}<-0.36$ for a kinematic study by using the $V-J$ versus $J-H$ diagram.
Their space motions and Galactic orbits were computed with the parallaxes and proper motions published by Gaia DR2 and radial velocities measured
from LAMOST DR5 spectra. He-rich stars with $\log(y)\ge0$ show diverse space motions and the largest standard deviations
of the space velocity components. We also presented a kinematic population classification of the four hot subdwarf helium groups based on
their positions in the $U-V$ velocity diagram, $J_{Z}-$eccentricity diagram and their Galactic orbits. Compared to the predictions of
the binary population synthesis by \cite{2008A&A...484L..31H}, the relative numbers of the four hot subdwarf helium groups in the halo, thin disk,
and thick disk suggest that He-deficient stars with $\log(y)<-2.2$ likely origin from the stable RLOF channel,  He-deficient stars
with $-2.2\le\log(y)<-1$ from the CE ejection channel, and He-rich stars with $\log(y)\ge0$ from the merger channel of
double He white dwarf stars. Comparing our results to the predictions of the population synthesis by \cite{2017ApJ...835..242Z}, we find that the channel of helium white dwarf mergers with low mass main sequence stars may not be the main contribution of the production of He-rich stars with $-1\le\log(y)<0$, especially in the thin disk.

\acknowledgments
The research presented here is supported by
the National Natural Science Foundation of China under
grant no. U1731111.  P.N. acknowledges support from the Grant Agency of the Czech Republic (GA\v{C}R 18-20083S).
Z.H. acknowledges support from the National Natural Science Foundation of China (11521303 and 11733008) and foundation of Yunnan province (2017HC018).
Guoshoujing Telescope (the Large Sky Area Multi-Object Fiber Spectroscopic
Telescope LAMOST) is a National Major Scientific
Project built by the Chinese Academy of Sciences.
Funding for the project has been provided by the National
Development and Reform Commission. LAMOST
is operated and managed by the National Astronomical
Observatories, Chinese Academy of Sciences.
This work has made use of data from the European Space Agency (ESA) mission
{\it Gaia} (\url{https://www.cosmos.esa.int/gaia}), processed by the {\it Gaia}
Data Processing and Analysis Consortium (DPAC,
\url{https://www.cosmos.esa.int/web/gaia/dpac/consortium}). Funding for the DPAC
has been provided by national institutions, in particular the institutions
participating in the {\it Gaia} Multilateral Agreement.
This research has used the services of \mbox{\url{www.Astroserver.org}}.
\software{lmfit-py (https://lmfit.github.io/lmfit-py/), \,astropy \citep{2013A&A...558A..33A, 2018AJ....156..123A}, \,TOPCAT (v4.6; \citealt{2005ASPC..347...29T,2018arXiv181109480T}), \,galpy \citep{2015ApJS..216...29B}, \,Tlusty (v205; \citealt{1995ApJ...439..875H}),\, Synspec (v51; \citealt{2011ascl.soft09022H})}.



\begin{longrotatetable}
\begin{deluxetable*}{llrrrrrrrrrrrrrrr}
\tablecaption{Parameters of 32 strong composite spectra Hot subdwarf stars selected from Gaia DR2 and LAMOST DR5. \label{tab:tab1}}
\tablewidth{1500pt}
\tabletypesize{\scriptsize}
\tablehead{
\colhead{LAMOST}       & \colhead{Name}    & \colhead{$G$}      & \colhead{$G_{\rm BP}-G_{\rm RP}$} & \colhead{$A_{G}$} & \colhead{$E(G_{\rm BP}-G_{\rm RP})$} &
\colhead{$parallax$}   & \colhead{$\pm$}   &  \colhead{$pmra$}   & \colhead{$\pm$} & \colhead{$pmdec$}  & \colhead{$\pm$}  & \colhead{$RV$}     & \colhead{$\pm$}\\
\colhead{}             & \colhead{}        & \colhead{$(mag)$}    & \colhead{$(mag)$}                    & \colhead{$(mag)$}  & \colhead{$(mag)$} &
\colhead{$(mas)$}      & \colhead{}        & \colhead{$(mas\,yr^{-1})$} & \colhead{} & \colhead{$(mas\,yr^{-1})$} &  \colhead{} & \colhead{$(km\,s^{-1})$} & \colhead{}
 }
\startdata
J004825.94+550008.5& --             &11.7782 & 0.2737 & 0.9442 & 0.4648 &11.2188 & 1.0215 & -3.996 & 1.012 & 19.819 & 1.166 &  -40 & 27\\
J005701.09+082524.3&PB6221          &16.5535 &-0.4228 & 0.1950 & 0.0651 & 0.2601 & 0.1193 & -0.012 & 0.196 & -5.497 & 0.107 & -122 & 32\\
J023254.08+370421.4&Pul-3190637     &15.8523 &-0.4364 & 0.1563 & 0.0440 & 0.7442 & 0.0780 & -5.308 & 0.174 & -5.816 & 0.136 &  -66 & 17\\
J030112.86+184054.1&PG0258+185      &15.4191 & 0.0773 & 0.4543 & 0.2057 & 0.6126 & 0.0631 &  8.611 & 0.120 & -4.728 & 0.110 &   36 & 31\\
J030754.31+152514.3&PG0305+152      &14.3838 &-0.2912 & 0.5349 & 0.2489 & 1.1201 & 0.0576 &  1.102 & 0.107 &  1.764 & 0.090 &   -1 & 11\\
J053505.93-041255.5&--              &16.4049 & 0.3528 & 0.8458 & 0.4134 & 0.6355 & 0.0784 & -2.619 & 0.142 & -6.230 & 0.124 &  -65 & 63\\
J055926.67+451916.9&--              &12.9139 &-0.0647 & 0.6283 & 0.2986 & 1.3919 & 0.0675 &  0.807 & 0.100 & -7.285 & 0.085 & -287 & 85\\
J062037.41+290352.2&TYC1890-1473-1  &11.8055 & 0.3466 & 0.9418 & 0.4635 & 4.0031 & 0.7814 & -8.683 & 1.340 &-10.024 & 1.220 &    9 & 21\\
J063144.96+190537.1&TYC1336-706-1   &10.8762 & 0.3257 & 0.9337 & 0.4593 & 6.7460 & 1.0332 & -3.168 & 1.672 &-14.177 & 1.485 &   20 & 26\\
J063418.32+230955.7&--              &12.9482 & 0.2515 & 0.5662 & 0.2656 & 3.7060 & 0.7548 &  2.346 & 1.376 & -2.570 & 1.211 &   35 & 22\\
J064643.21+120929.9&--              &12.7020 & 0.3619 & 0.8402 & 0.4105 & 4.7279 & 0.5690 &  0.052 & 0.901 & -3.746 & 0.800 &   11 & 23\\
J065314.44+054733.7&TYC161-49-1     &11.1375 & 0.3661 & 0.9700 & 0.4781 & 8.2284 & 0.5884 & -2.536 & 0.958 & -0.934 & 0.855 &   10 & 30\\
J065930.72+055227.5&--              &17.2799 &-0.0851 & 0.7712 & 0.3742 & 0.1713 & 0.1037 & -1.735 & 0.196 & -1.045 & 0.202 &   47 & 23\\
J074852.07+454903.7&2MASSJ07485208+4549037   &16.0743 &-0.3684 & 0.1601 & 0.0460 & 0.3858 & 0.0720 &-4.166 & 0.107  &-4.947 & 0.089 &   93 & 28\\
J080259.80+411437.9&KUV07596+4123   &15.2668 &-0.3912 & 0.1495 & 0.0402 & 0.3441 & 0.0651 & -4.000 & 0.092 & -0.903 & 0.063 &   -2 & 11\\
J081339.72+562709.0&SBSS0809+566    &14.0125 &-0.3921 & 0.1383 & 0.0341 & 1.2534 & 0.0567 & -5.977 & 0.075 & -3.762 & 0.059 &   -47 & 25\\
J084844.73+124147.9&PG0845+129      &16.5685 &-0.3226 & 0.0646 &-0.0064 & 0.2044 & 0.0877 &  7.975 & 0.149 &-20.693 & 0.113 &   -1 & 10\\
J093920.67+032632.2&PG0936+037      &15.4625 &-0.4161 & 0.1010 & 0.0136 & 0.5534 & 0.0516 & -8.822 & 0.087 & -1.601 & 0.076 &  -13 & 25\\
J094623.10+040456.0&PG0943+043      &15.7064 &-0.4520 & 0.1125 & 0.0199 & 0.6364 & 0.1001 & -7.687 & 0.130 & -4.389 & 0.115 &   57 & 31\\
J095101.34+034757.3&PG0948+041      &15.8887 &-0.2599 & 0.1052 & 0.0159 & 0.4064 & 0.0981 &-11.717 & 0.123 & -8.057 & 0.111 &  125 & 24\\
J103853.99+525847.5&PG1035+532      &16.1848 &-0.4484 & 0.0535 &-0.0126 & 0.5059 & 0.0886 & -6.032 & 0.103 & -3.347 & 0.162 &   81 & 28\\
J111829.83+271703.3&Ton577          &15.6811 &-0.5861 & 0.0492 &-0.0149 & 0.8815 & 0.0783 &  2.568 & 0.122 &-24.423 & 0.105 &   96 & 21\\
J111904.87+295153.5&Ton63           &14.3048 &-0.4717 & 0.0408 &-0.0196 & 1.2154 & 0.0779 &-14.928 & 0.139 &-12.116 & 0.231 &    1 & 28\\
J125004.42+550602.1&GD319           &12.6898 &-0.4782 & 0.0336 &-0.0236 & 4.0815 & 0.0555 &-66.921 & 0.080 &-12.317 & 0.078 &   19 & 24\\
J131333.00+365649.0&HZ40            &14.5488 &-0.4734 & 0.0300 &-0.0256 & 0.8978 & 0.0520 & -5.238 & 0.071 & -4.831 & 0.072 &    5 & 28\\
J135401.84+134350.5&PG1351+139      &15.9826 &-0.1468 & 0.0669 &-0.0052 & 0.5990 & 0.0845 & -6.800 & 0.151 &-12.103 & 0.110 &   -4 & 26\\
J144933.64+244336.2&LB731           &15.7003 &-0.4741 & 0.0930 & 0.0092 & 0.3143 & 0.0782 & -9.481 & 0.121 & -9.436 & 0.143 & -123 & 31\\
J162936.86+312411.9&SDSSJ162936.86+312412.0&18.5109 &-0.6316 & 0.0632 &-0.0072 & 0.0783 & 0.1747 & -4.024 & 0.284 & -2.908 & 0.331 & -155 & 23\\
J165809.14+214046.4&SDSSJ165809.15+214046.4&16.0656 &-0.4268 & 0.1673 & 0.0500 & 0.5990 & 0.0586 & -2.693 & 0.076 & -2.018 & 0.095 &  -20 & 26\\
J171952.09+000550.5&--             &16.1443& 0.1223 & 0.9641 & 0.4751 & 0.9249 & 0.0682 &  2.976 & 0.116 & -2.586 & 0.092 &  -40 & 32\\
J212728.98+311841.1&--             &14.4421& 0.1057 & 0.5285 & 0.2454 & 0.8926 & 0.0947 & 11.702 & 0.216 &  6.336 & 0.220 &  -71 & 27\\
J235745.40+242622.5&PG2355+242     &16.3563& -0.3013 & 0.369 & 0.1597 & 0.3716 & 0.1037 & -1.924 & 0.252 & -8.487 & 0.083 &  -15 & 25\\
\enddata
\end{deluxetable*}
\end{longrotatetable}

\begin{deluxetable*}{rlll}
\tablecaption{Atmospheric parameters and space positions for 892 non-composite spectra hot subdwarf stars selected from Gaia DR2 and LAMOST DR5.\label{tab:tab2}}
\tablewidth{0pt}
\tablehead{
\colhead{Num} & \colhead{Label}    & \colhead{Explanations}
}
\startdata
  1  & LAMOST              & LAMOST target\\
  2  & NAME                & Target name in the SIMBAD Astronomical Database\\
  3  & $RA$                & Barycentric right ascension (ICRS) at Ep=2000.0\\
  4  & $DEC$               & Barycentric declination (ICRS) at Ep=2000.0\\
  5  & $G$                 & G-band mean magnitude\\
  6  & $G_{BP}-G_{RP}$     & $G_{BP}-G_{RP}$ colour\\
  7  & $AG$                & Estimate of extinction in the G band\\
  8  & $E(G_{BP}-G_{RP})$  & Estimate of redenning $G_{BP}-G_{RP}$ colour\\
  9  & $V-J$               & Vmag-Jmag colour\\
  10 & $J-H$               & Jmag-Hmag colour\\
  11 & $parallax$          & Absolute stellar parallax\\
  12 & $e\_parallax$       & Standard error of parallax\\
  13 & $pmra$              & Proper motion in right ascension direction\\
  14 & $e\_pmra$           & Standard error of proper motion in right ascension direction\\
  15 & $pmdec$             & Proper motion in declination direction\\
  16 & $e\_pmdec$          & Standard error of proper motion in declination direction\\
  17 & $RV$                & Radial velocity from LAMOST spectra\\
  18 & $e\_RV$             & Standard error of Radial velocity\\
  19 & $T_{\rm eff}$       & Stellar effective temperature\\
  20 & $e\_T_{\rm eff}$    & Standard error of Stellar effective temperature\\
  21 & $\log\,g$           & Stellar surface gravity\\
  22 & $e\_\log\,g$        & Standard error of Stellar surface gravity\\
  23 & $log(y)$            & Stellar surface He abundance y = n(He)/n(H)\\
  24 & $e\_\log(y)$        & Standard error of Stellar surface He abundance\\
  25 & $X$                 & Galactic position towards the Galactic centre\\
  26 & $e\_X$              & Standard error of Galactic position towards the Galactic centre\\
  27 & $Y$                 & Galactic position in the direction of Galactic rotation\\
  28 & $e\_Y$              & Standard error of Galactic position in the direction of Galactic rotation\\
  29 & $Z$                 & Galactic position in the direction of the north Galactic pole\\
  30 & $e\_Z$              & Standard error of Galactic position in the direction of the north Galactic pole\\
 \enddata
 \tablecomments{Only a portion of this table is shown here to demonstrate its form and content. Machine-readable version of the full table are available on \url{https://zenodo.org/record/3136102\#.XOUPGPnE8l8}.}
\end{deluxetable*}

\begin{deluxetable*}{rlll}
\tablecaption{Orbital parameters, galactic velocities for 747 hot subdwarf stars selected from Gaia DR2 and LAMOST DR5. \label{tab:tab3}}
\tablewidth{0pt}
\tablehead{
\colhead{Num} & \colhead{Label}    & \colhead{Explanations}
}
\startdata
  1  & LAMOST              & LAMOST target\\
  2  & NAME                & Target name in the SIMBAD Astronomical Database\\
  3  & $RA$                & Barycentric right ascension (ICRS) at Ep=2000.0\\
  4  & $DEC$               & Barycentric declination (ICRS) at Ep=2000.0\\
  5  & $parallax$          & Absolute stellar parallax\\
  6  & $e\_parallax$       & Standard error of parallax\\
  7  & $pmra$              & Proper motion in right ascension direction\\
  8  & $e\_pmra$           & Standard error of proper motion in right ascension direction\\
  9  & $pmdec$             & Proper motion in declination direction\\
  10 & $e\_pmdec$          & Standard error of proper motion in declination direction\\
  11 & $RV$                & Radial velocity from LAMOST spectra\\
  12 & $e\_RV$             & Standard error of Radial velocity\\
  13 & $U$                 & Galactic radial velocity positive towards the Galactic centre\\
  14 & $e\_U$              & Standard error of Galactic radial velocity\\
  15 & $V$                 & Galactic rotational velocity in the direction of Galactic rotation\\
  16 & $e\_V$              & Standard error of Galactic rotational velocity\\
  17 & $W$                 & Galactic velocity in the direction of the north Galactic pole\\
  18 & $e\_W$              & Standard error of Galactic velocity in the direction of the north Galactic pole\\
  19 & $Rap$               & Apocenter radius from the numerical orbit integration\\
  20 & $e\_Rap$            & Standard error of the apocenter radius\\
  21 & $Rperi$             & Pericenter radius from the numerical orbit integration\\
  22 & $e\_Rperi$          & Standard error of the pericenter radius\\
  23 & $z_{\rm max}$       & Maximum vertical height from the numerical orbit integration\\
  24 & $e\_z_{\rm max}$    & Standard error of the maximum vertical height\\
  25 & $ec$                & Eccentricity from the numerical orbit integration\\
  26 & $e\_ec$             & Standard error of the eccentricity\\
  27 & $J_{z}$             & Z-component of angular momentum\\
  28 & $e\_J_{z}$          & Standard error of the z-component of angular momentum\\
  29 & $z_{n}$             & Normalised z-extent of the orbit\\
  30 & $e\_z_{n}$          & Standard error of the normalised z-extent of the orbit\\
  31 & pops                & Population classification: Halo (H), thick disk (TK), thin disk (TH)\\
 \enddata
 \tablecomments{Only a portion of this table is shown here to demonstrate its form and content. Machine-readable version of the full table are available on \url{https://zenodo.org/record/3136102\#.XOUPGPnE8l8}.}
 \end{deluxetable*}

\begin{deluxetable*}{lrrrrrrrr}
\tablecaption{Mean values and standard deviations of the Galactic velocities for the hot subdwarf helium groups. \label{tab:tab4}}
\tablewidth{0pt}
\tabletypesize{\scriptsize}
\tablehead{
\colhead{Subsample} & \colhead{N} &
\colhead{$\bar{U}$} & \colhead{$\sigma_{U}$} &
\colhead{$\bar{V}$} & \colhead{$\sigma_{V}$} &
\colhead{$\bar{W}$} & \colhead{$\sigma_{W}$}
}
\startdata
All stars             & 747  &  7  &  69 & 189 & 77  & -2  & 47\\
$\log(y)\ge0$         & 124  &  6  &  90 & 155 & 105 & -10 & 56\\
$-1\le\log(y)<0$      & 43   &  4  &  80 & 188 & 75  & -5  & 44 \\
$-2.2\le\log(y)<-1$   & 216  &  5  &  62 & 198 & 63  & 1   & 39 \\
$\log(y)<-2.2$        & 364  &  8  &  63 & 196 & 70  & -1  & 47  \\
\enddata
\end{deluxetable*}

\begin{deluxetable*}{lrrrrrrrrrrr}
\tablecaption{Mean values and standard deviations of the Galactic orbital parameters: eccentricity ($e$), normalised z-extent ($z_{\rm n}$),  maximum vertical amplitude ($z_{\rm max}$), apocentre ($R_{\rm ap}$) and pericentre  ($R_{\rm peri}$) for the hot subdwarf helium groups.\label{tab:tab5}}
\tablewidth{0pt}
\tablehead{
\colhead{Subsample}           & \colhead{N}                      &
\colhead{$\overline{e}$}           & \colhead{$\sigma_{e}$}           &
\colhead{$\overline{z_{\rm n}}$}   & \colhead{$\sigma_{z_{\rm n}}$}   &
\colhead{$\overline{z_{\rm max}}$} & \colhead{$\sigma_{z_{\rm max}}$} &
\colhead{$\overline{R_{ap}}$}      & \colhead{$\sigma_{R_{ap}}$}       &
\colhead{$\overline{R_{peri}}$}    & \colhead{$\sigma_{R_{peri}}$}
}
\startdata
All stars           & 747  &  0.29  &  0.20 & 0.23 & 0.45 & 2.05 & 2.96 & 10.68 &  3.72  & 6.06 & 2.53 \\
$\log(y)\ge0$       & 124  &  0.36  &  0.25 & 0.36 & 0.56 & 3.25 & 4.07 & 11.40 &  5.91  & 5.58 & 3.35 \\
$-1\le\log(y)<0$    & 43   &  0.33  &  0.27 & 0.26 & 0.47 & 2.11 & 3.16 & 10.33 &  2.68  & 5.60 & 2.85 \\
$-2.2\le\log(y)<-1$ & 216  &  0.26  &  0.19 & 0.21 & 0.54 & 1.64 & 2.00 & 10.43 &  2.78  & 6.16 & 2.19 \\
$\log(y)<-2.2$      & 364  &  0.27  &  0.17 & 0.19 & 0.31 & 1.87 & 2.85 & 10.63 &  3.30  & 6.22 & 2.32 \\
\enddata
\end{deluxetable*}

\begin{deluxetable*}{lclclclclc}
\tablecaption{Population classification for hot subdwarf stars selected from Gaia DR2 and LAMOST DR5.\label{tab:tab6}}
\tablewidth{0pt}
\tablehead{
\colhead{Subsample}  &
\colhead{N}          &
\colhead{Thin Disk}  &
\colhead{Thick Disk} &
\colhead{Halo}
}
\startdata
All stars           & 747  &  328 & 312 & 107  \\
$\log(y)\ge0$       & 124  &  32  &  57 & 35  \\
$-1\le\log(y)<0$    & 43   &  22  &  14 & 7  \\
$-2.2\le\log(y)<-1$ & 216  & 112  &   82 & 22  \\
$\log(y)<-2.2$      & 364  & 162  &  159 & 43  \\
\enddata
\end{deluxetable*}

\end{document}